\documentclass[pra,aps,10pt,nofootinbib,twocolumn,superscriptaddress]{revtex4-1}

\def\Title{Stability and decay of Bloch oscillations in presence of 
time-dependent nonlinearity}

\usepackage[colorlinks,pdftitle=\Title]{hyperref}
\usepackage{amsmath}  % e.g. for \eqref{}
\usepackage{amsfonts} % e.g. for \mathbb{N}
\usepackage[utf8]{inputenc}
\usepackage[pdftex]{graphicx}
\usepackage{subfigure}
\usepackage{bm}       % bold math

\usepackage[normalem]{ulem}

\usepackage{enumitem}
\setitemize[0]{topsep=2pt, itemsep=0pt, parsep=2pt, leftmargin=1.5em}

\newcommand{\Heff}{H_\text{cc}}
\newcommand{\omegaB}{\omega_\text{B}}
\newcommand{\TB}{T_\text{B}}
\newcommand{\xB}{x_\text{B}}
\newcommand{\mv}[1]{\left\langle#1\right\rangle}
\renewcommand{\Re}{\mathrm{Re}}

\newcommand{\rmd}{\mathrm{d}}

\newcommand{\tavg}[1]{\overline{#1}}		% time-average (coarse graining)

\newcommand{\matr}[4]{\left(\begin{array}{cc}#1&#2\\#3&#4\end{array}\right)}   % 2x2 Matrix
\newcommand{\cvect}[2]{\left(\begin{array}{c}#1\\#2\end{array}\right)}         % Column vector
\newcommand{\Efr}{\omega_{k}^0}	% Free kinetic energy
\newcommand{\EBg}{\omega_{k}}	% Bogoliubov energy

\begin{document}
\title{\Title}
\author{Christopher Gaul}
\affiliation{GISC, Departamento de F\'{\i}sica de Materiales, Universidad Complutense, E-28040 Madrid, Spain}

\author{Elena D\'{\i}az} 
\affiliation{Institute for Materials Science, Technische Universit\"{a}t Dresden, D-01062 Dresden, Germany} 
\affiliation{GISC, Departamento de F\'{\i}sica de Materiales, Universidad Complutense, E-28040 Madrid, Spain}

\author{Rodrigo P. A. Lima}
\affiliation{Instituto de F\'{\i}sica, Universidade Federal de Alagoas, Macei\'{o} AL 57072-970, Brazil}
\affiliation{GISC, Departamento de F\'{\i}sica de Materiales, Universidad Complutense, E-28040 Madrid, Spain}

\author{Francisco Dom\'{\i}nguez-Adame}
\affiliation{GISC, Departamento de F\'{\i}sica de Materiales, Universidad Complutense, E-28040 Madrid, Spain}

\author{Cord A. M\"{u}ller}
\affiliation{Centre for Quantum Technologies, National University of Singapore, Singapore 117543, Singapore}

\date{\today}

\begin{abstract}
We consider Bloch oscillations of Bose-Einstein condensates in
presence of a time-modulated $s$-wave scattering length. Generically,
interaction leads to dephasing and decay of the wave packet. 
Based on a cyclic-time argument, we find---additionally to the linear Bloch oscillation and a rigid soliton
solution---an infinite family of modulations that lead to a periodic time evolution of the wave packet.
In order to quantitatively describe the dynamics of Bloch oscillations in presence of time-modulated interactions, 
we employ two complementary methods: collective-coordinates and the linear stability analysis of an extended wave packet.
We provide instructive examples and address the question of robustness against external perturbations.
\end{abstract}

\maketitle

%----------------------------------------------------
\section{Introduction}\label{secIntroduction}
Quantum dynamics of atomic Bose-Einstein condensates~(BECs) in optical lattices,
formed by counter-propagating laser
beams~\cite{Anderson1998,Cataliotti2001,Morsch2001}, bears much resemblance to
electron dynamics in solid-state crystals. 
For this reason, and due to the vanishingly small contribution of decoherence effects, BECs are 
an ideal test ground for quantum transport of matter waves in complex
environments.  
Perhaps one of the best examples 
are Bloch oscillations~(BOs), a phenomenon predicted by
Zener~\cite{Zener1934} 
based on the band-structure framework established by Bloch \cite{Bloch1929}:
quantum particles in periodic potentials subjected to a constant
force do not accelerate uniformly in real space, but oscillate instead.
Because of defects and decoherence, BOs cannot be observed in conventional solids.
With the mastery of ultracold atomic gases, however, BOs have been observed as a 
periodic motion of ensembles of ultracold atoms~\cite{Ben1996,Wilkinson1996} and BECs~\cite{Anderson1998,Cataliotti2001,Morsch2001} 
in tilted optical lattices. 

The basic phenomenon of BOs can be well understood within a semi-classical
framework. Let us consider a wave packet with a narrow momentum distribution in
a lattice. If the wave packet is accelerated by a constant force $-F$ (like
gravity, for massive particles, or an electric field, for charged ones), then its momentum $\hbar k
= -F t$ will increase linearly.
In a periodic potential with spatial period $d$, the dispersion relation of free particles is replaced with a
band-structure dispersion $\epsilon_{n k}$ with band index $n$ and
quasi-momentum $k$. 
In the tight-binding description, which is appropriate for a very deep lattice, 
the lowest-band dispersion reads $\epsilon_k \propto [1-\cos(k d)]$. 
Now, quasi-momentum and velocity are no longer proportional. Instead, a wave
packet that is uniformly accelerated across the Brillouin zone has the group velocity  
$v_\mathrm{g} \propto \partial_k \epsilon(k) \propto \sin(k d) =
-\sin(\omegaB t)$, oscillating with the Bloch frequency $\omegaB=F d/\hbar$.
Consequently, the wave packet oscillates back and forth in real space.
Also, related coherent phenomena have been realized with ultracold atoms.
For instance, when the external force is modulated harmonically in time, the interwell tunneling of a BEC can be suppressed \cite{Lignier2007},
as predicted theoretically in Ref.~\cite{Eckardt2005}.
In addition, the simultaneous action of both constant and time-harmonic forces may lead to giant matter-wave oscillations called super-BOs due to the beating of the usual BOs and the drive \cite{Haller2010}.

Because BOs rely on the coherent reflection of waves, they are very sensitive to any kind
of dephasing generated by interaction effects or lattice imperfections. 
Any deviation from perfect periodicity causes random 
scattering of different $k$-components of the wave packet. Thus, its momentum
distribution starts to broaden and the coherent oscillations in real space are
destroyed. This is the situation in crystalline solids, where the lattice
spacing $d$ is given by atomic distances, which are so short that electrons
suffer from scattering events long before their quasi-momentum reaches
the Brillouin-zone edge $\pi/d$. One way to overcome this problem
is to artificially increase the lattice spacing, thus shortening the Bloch period, as achieved in
semiconductor superlattices~\cite{Feldmann1992,Leo1992}. 

Experiments with ultracold dilute gases offer very clean experimental conditions and open new possibilities.
Atomic length scales are replaced by optical length scales.
Atom-atom interactions---the main source of dephasing---are dominated by $s$-wave scattering.
Many alkali species (e.g.\ ${}^7$Li \cite{Khaykovich2002}, ${}^{133}$Cs \cite{Tiesinga1993,Gustavsson2008}) allow tuning the $s$-wave scattering length by means of a Feshbach resonance~\cite{Tiesinga1993,Timmermans1999,Koehler2006,Chin2010}.
The $s$-wave scattering length can be tuned in a wide range, including a smooth crossover to negative values, i.e., attractive interaction.
By suppressing the interaction entirely, one can observe very long-living BOs (up to $10^4$ cycles in Ref.~\cite{Gustavsson2008}).
There are always residual experimental uncertainties, 
like a fluctuating $s$-wave scattering length, 
whose effect should be considered. Moreover, the scattering length can
be deliberately modulated in time, which opens a pathway to new
effects and interesting spectroscopic applications. 
For example, a harmonic modulation in time can be used to
probe the collective excitations of trapped BEC
\cite{Pollack2010,Vidanovic2011}. 

In this work we present a detailed study of the stability and decay of BOs in tilted optical lattices for BECs with an atom-atom contact interaction that is modulated harmonically in time.
Throughout the article, we discuss all results in the BEC context. But we like to emphasize from the outset that 
our analysis is based on mathematical properties of the nonlinear Schr\"odinger equation and thus applies to 
all physical systems governed by equation \eqref{eqTightBinding} below. In particular, a very clean realization 
is provided by 1D lattices of optical wave guides 
\cite{Morandotti1999,Pertsch1999,Sapienza2003}. 
In previous work \cite{Gaul2009,Diaz2010}, we have identified an infinite family of harmonic modulations $g(t)$ that guarantee long-living BOs on the mean-field level.
We have studied both the stable and unstable cases using, respectively, a collective-coordinates (CC) approach \cite{Trombettoni2001} as well as a linear stability analysis within Floquet theory \cite{Markley2004}.
In this article, we re-derive these results in greater detail and extend them in several important aspects.

The paper is organized as follows. In Sec.~\ref{secModel} we introduce the
tight-binding approximation to the Gross-Pitaevskii description, suitable for BECs in deep optical lattices.
A numerical solution of the discrete Gross-Pitaevskii equation with time-harmonic atom-atom interaction shows the occurrence of stable and unstable dynamics of the BEC, depending on the frequency and phase of the modulation.
In Sec.~\ref{secSymmetry} we prove, within the smooth-envelope approximation, the existence of an infinite family of interactions leading to stable BOs, which is at odds with a quasistatic soliton stability criterion. 
We generalize the cyclic-time argument
developed in Ref.\ \cite{Diaz2010} to cover all the solutions found by a
different method in \cite{Gaul2009}, and derive the limit of validity of the
wide-envelope ansatz.
Next, Sec.~\ref{secCollectiveCoordinates} recapitulates the CC approach of Ref.\ \cite{Gaul2009} and improves the physical interpretation in terms of the momentum variance. 
The impact of several relevant modulations of the interaction is discussed in detail.
The CC approach captures satisfactorily the effects of time-dependent atom-atom interaction, as long as the wave-packet shape is essentially preserved. 
The decay of BOs under unstable modulations is described in
Sec.~\ref{secStability}, where we develop a linear stability analysis of wide
wave packets. Via perturbative Floquet theory, we study the growth of
perturbations that ultimately destroy the wave packet, covering a wider class of
unstable perturbations than presented in Ref.\ \cite{Diaz2010}. 
After discussing the respective regimes of
validity of our approaches, we summarize and conclude the article in Sec.~\ref{secConclusions}.

%----------------------------------------------------
\section{Mean-field tight-binding model}\label{secModel}
In typical BEC experiments, atomic gases are very dilute, in the sense
that the interparticle spacing exceeds the $s$-wave scattering length,
which allows for a description within Gross-Pitaevskii theory
\cite{Dalfovo1999}. 
The condensate is initially created in a harmonic trap and then loaded into an optical lattice potential $V({\bm r})$ with transverse confinement \cite{Gustavsson2008}.
The condensate amplitude $\Psi(\bm{r},t)$ then evolves according to the Gross-Pitaevskii equation
\begin{eqnarray}
\label{eqGPE_BO}
i \hbar \,\frac{\partial}{\partial t} \Psi ({\bm r},t)
& = & \left[-\frac{\hbar^2}{2m}\,\nabla^2 + V ({\bm r}) + F z
\right] \Psi({\bm r},t) \nonumber \\
 & & +  g_{\rm 3D} |\Psi({\bm r},t)|^2 \Psi({\bm r},t)\ .
\end{eqnarray}
The homogeneous force $F$ describes a uniform acceleration, for instance by gravity, along an axis, which we take to be the $z$ axis. 
Even in absence of the force $F$, the initial wave packet is not the
ground-state configuration, and the condensate tends to spread across
the lattice. 
In the case of repulsive self-interaction $g_{\rm 3D} = 4\pi \hbar^2
a_s /m >0$  ($a_s$ is the $s$-wave scattering length), this tendency
is enhanced. 
In the opposite case $a_s<0$, self-attraction counteracts dispersion and allows for soliton solutions \cite{Khaykovich2002}.

Since the phenomenon of BO takes place only in the longitudinal $z$ direction, 
we consider a lattice potential with strong transverse confinement, such that 
the transverse degrees of freedom remain frozen in their harmonic-oscillator ground state.
In particular, we exclude from our study the regime of weak transverse confinement that results in stacks of pancake-shaped BECs. These are
  prone to transverse excitations in the presence of the time-dependent,
  sign-changing nonlinearity we consider below. 
For an extensive study of the excitation of transverse degrees of freedom, see Ref.\ \cite{Marzlin2005}.
Integrating out the transverse degrees of freedom then results in the one-dimensional Gross-Pitaevskii equation
\begin{eqnarray}
\label{eqGPE1D}
i \hbar \, \frac{\partial}{{\partial}t}\, \Psi_{\rm 1D}(z,t) 
&=& \left[-\frac{\hbar^2}{2m} \partial_z^2 + V_{\rm 1D}(z) + F z\right] \Psi_{\rm 1D}(z,t) 
\nonumber\\
&&+ g_{\rm 1D} \, |\Psi_{\rm 1D}(z,t)|^2 \Psi_{\rm 1D}(z,t) \ ,
\end{eqnarray}
with the effective interaction parameter $g_{\rm 1D} = m \omega_\perp g_{\rm 3D}/2\pi\hbar$ \cite{Morsch2006}. 
In order for Eq.\ \eqref{eqGPE1D} to be valid, the transverse oscillator energy $\hbar \omega_\perp$ must exceed all other relevant energies.

If the potential $V_{\rm 1D}(z)= V_0 \sin^2(\pi z / d)$,
with lattice constant $d$, amplitude $V_0>0$, 
and local oscillator frequency 
$\omega_\parallel = \pi \sqrt{2V_0/m}/d$, 
is sufficiently deep, only the local harmonic oscillator ground state
(or Wannier function of the lowest band) of each lattice well is populated. 
The condensate is represented by the complex amplitudes $\Psi_n$
for occupying the lattice wells 
centered at $z_n= n d$. 
A tight-binding equation of motion is found by integrating also over the $z$ coordinate:
\begin{equation}
\label{eqTightBinding1}
i \hbar \dot \Psi_n = -J(\Psi_{n+1}+\Psi_{n-1}) + 
F d n  \Psi_{n} + g  |\Psi_{n}|^2 \Psi_{n} \ .
\end{equation}
Neighboring sites are coupled by the tunneling matrix element  
$J \approx 4\pi^{-1/2}E_{\rm r} \left({V_0}/{E_{\rm r}}\right)^{3/4} \exp(-2\sqrt{V_0/E_{\rm r}})$,
where $E_{\rm r} = \hbar^2 \pi^2/(2m d^2)$ is the recoil energy~\cite{Morsch2006}. 
The tight-binding interaction parameter 
$g = N \sqrt{{m \omega_\parallel}/{2\pi \hbar}} \, g_{\rm 1D}$
contains the total number of particles $N$ because we choose to normalize the discrete wave function as $\sum_n |\Psi_n|^2 =1$.
Eq.\ \eqref{eqTightBinding1} is valid only for very deep 
transverse and longitudinal trapping potentials, for which
$\hbar \omega_\parallel, \hbar \omega_\perp  \gg  \|\mu\|_\infty $, 
where $\|\mu \|_\infty = \max_n |g\Psi_n^2|$ is the maximum local mean-field interaction energy.  
Under these conditions the shape of the local wave functions
does not depend much on the occupation \cite{Smerzi2003,Trombettoni2001},
and the tight-binding parameters $J$ and $g$ are also not affected.

\begin{figure}
\begin{center}
\includegraphics[angle=270,width=0.48\linewidth]{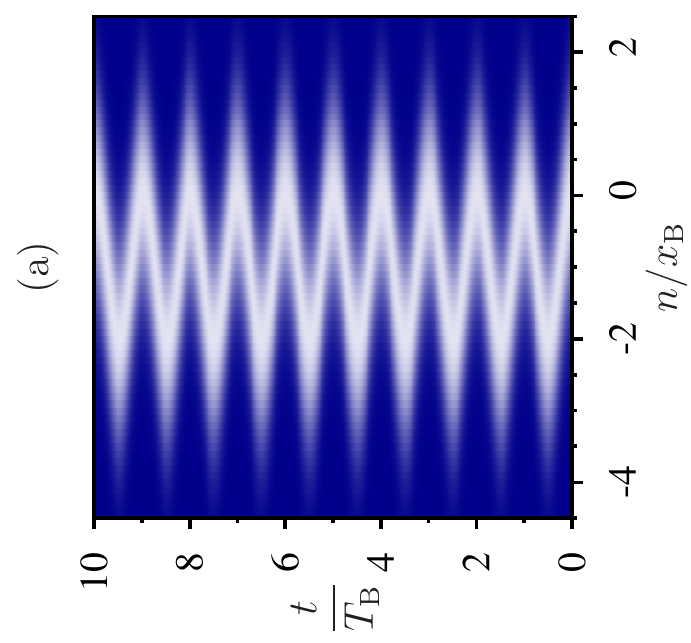}\hfill
\includegraphics[angle=270,width=0.48\linewidth]{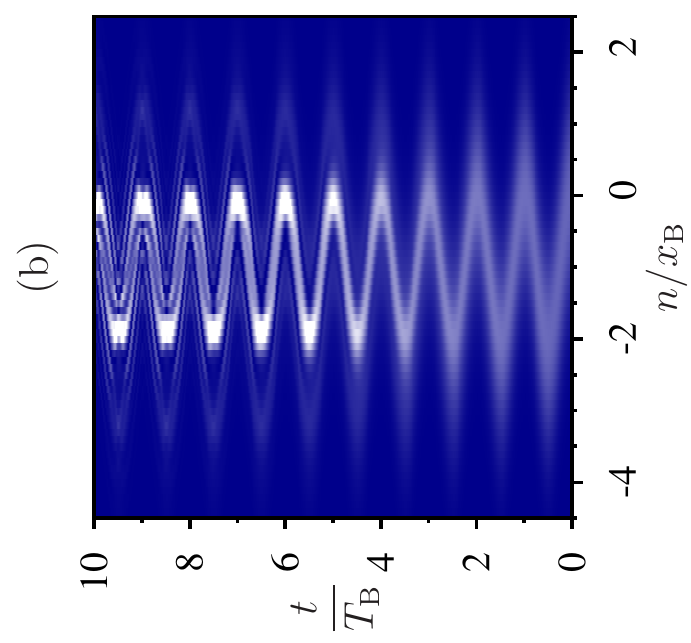}
\end{center}
\caption{(Color online) Condensate density $|\Psi_n(t)|^2$ as a
  function of position and time, obtained by numerical integration of
  Eq.~\eqref{eqTightBinding}. 
  Initially, the wave packet is at rest and has the Gaussian shape \eqref{Gaussian} with width $\sigma_0= 10$ chosen equal to the amplitude $\xB=2/F$, $F=0.2$ of free BOs. Two different harmonically modulated nonlinearities 
(a)~$g(t)=g_{\rm 0}\cos(F t)$ and (b)~$g(t)=g_{\rm 0} \sin(F t)$ with $g_{\rm 0}=1$ 
result in (a) stable and (b) unstable oscillations, respectively. 
}
\label{fig1}
\end{figure}

The tight-binding description~\eqref{eqTightBinding1} is equivalent to
a single-band description and thus neglects Landau-Zener tunneling
(LZT) to higher bands. 
Let us briefly discuss the conditions under which this approximation
is valid.
In the linear case ($g=0$), LZT can be neglected if the band gap $E_{\rm gap}=V_0/2$ \cite{Morsch2006,Mendez1991} is 
large enough, more precisely, if \cite{Niu1996}
\begin{align}\label{eqLZsuppr}
F d \, E_{\rm r} \ll E_{\rm gap}^2 \ .
\end{align}
In the case of constant interaction, the effective lattice potential is rescaled by a factor
$(1+4 E_{\rm int}/E_{\rm r})^{-1}$ \cite{Morsch2001}, reducing the
band gap by the same factor.
A typical experimental value of $V_0=4E_{\rm r}$ results in 
$E_{\rm r}/J = 8.55 \gg E_{\rm int}/J = g |\Psi_n|^2 \lesssim 0.1$ 
for typical parameters chosen below (see, e.g., Fig.\ \ref{fig1}).
Thus, we find that the relative correction to the lattice potential $E_{\rm int}/E_{\rm r}$ is smaller than 1.2\%, which does not change the previous validity condition \eqref{eqLZsuppr}.
Finally, in the case of modulated interaction, e.g., $g(t) = g_0
\cos(\omega t + \phi)$, 
one first has to ensure that the interaction energy remains smaller than the gap.
For the same parameters, this results in $E_{\rm int}/J \ll E_{\rm gap}/J = 17.1$, which is well fulfilled.
Second, the modulation frequency should not become resonant with the gap.
Below, we consider frequencies $\omega$ of the same order as the Bloch frequency $F d/\hbar$. Thus, with $V_0 / E_{\rm r}$ of order one, the condition $\hbar \omega \ll E_{\rm gap}$ is already included in Eq.~\eqref{eqLZsuppr}, which finally turns out to be the relevant condition 
also in the nonlinear cases.
We conclude that LZT can be neglected in the situations considered for this work, and that the single-band description
\eqref{eqTightBinding1} is justified.  

Hereafter we take the lattice constant $d$ and tunneling $J$ as units of length and energy, respectively, and set $\hbar=1$.
Eq.~\eqref{eqTightBinding1} then takes the form
\begin{equation}
\label{eqTightBinding}
i\dot\Psi_n = -\Psi_{n+1}-\Psi_{n-1} + F n\Psi_{n} + g(t)|\Psi_{n}|^2\Psi_{n} \ .
\end{equation}
In the noninteracting case $g = 0$, the atomic cloud oscillates with the Bloch
frequency $\omegaB=2\pi/\TB=F$ and amplitude $x_B \approx 2/F$ (in the
chosen units), as we recall in Sec.~\ref{linearBO} below. 
A constant nonlinearity $g\neq 0$ is known to rapidly dephase Bloch oscillations
\cite{Morandotti1999,Trombettoni2001,Witthaut2005,Gustavsson2008}. 
Our notation $g(t)$ emphasizes that we are interested in the effects
of an interaction that is modulated in time. In cold atomic
gases, this can be realized by means of an external
 magnetic field close to an appropriate Feshbach resonance
\cite{Tiesinga1993,Timmermans1999,Koehler2006,Chin2010}.
In arrays of nonlinear optical wave-guides, which also allow for a tight-binding
description like \eqref{eqTightBinding}, time is equivalent to the
propagation distance along the wave guides, and $g(t)$ could be
realized as a spatially modulated cubic nonlinearity \cite{Agrawal2007,Morandotti1999}.

In order to provide a foretaste of the interesting effects such a
time-dependent interaction can have, we solve Eq.~\eqref{eqTightBinding} numerically by means
of the fourth-order Runge-Kutta method for a wave packet with initial
Gaussian shape 
\begin{equation}\label{Gaussian} 
\Psi_n(0)=(2\pi\sigma_{0}^2)^{-1/4}\,\exp\left(-n^2/4\sigma_{0}^2\right)\ .
\end{equation} 
Figure~\ref{fig1} shows the condensate density $|\Psi_n(t)|^2$ 
as a function of position and time for two different harmonic
interactions: (a) $g(t)=g_0\cos(F t)$ and (b) $g(t)=g_0\sin(F t)$. 
In the first case (a), the Gaussian
shape is preserved over time, and the
interacting condensate performs long-living BOs with frequency $\omega_B=F$.  In the second case (b)
the initial shape is destroyed after a few cycles, as the wave function
develops satellite peaks, and BOs are rapidly destroyed.
 
%-------------------
%-------------------

These two cases differ solely by the relative phase between the interaction
modulation and the linear Bloch oscillation, which defines a reference time starting at $t=0$ 
when the wave packet is at rest. Strikingly, in the stable case (a), the
strongest repulsion $g=+g_0$ coincides with the upper turning point
of the wave packet, i.e.\  when the momentum is at the center of the
Brillouin zone and the positive mass disperses the wave packet. The
strongest attraction $g=-g_0$ occurs at the lower turning point, i.e.\
with the momentum at the Brillouin zone edge and negative mass
contracting the wave packet. Therefore, the observed stability clearly contradicts the simple quasistatic criterion, according to
which stable BOs should occur when the nonlinearity compensates the lattice
dispersion, instead of adding to it \cite{Salerno2008}. 

The results shown in Fig.~\ref{fig1} prompt at least the following questions, for which
we will provide the answers: (i) Which are the periodic
modulations $g(t)$ that lead to stable BOs? In the upcoming
Sec.~\ref{secSymmetry}, we use symmetry considerations to identify a family of stable modulations. 
(ii) How does the interaction affect the shape of an oscillating wave packet, both in stable and unstable cases?
Section~\ref{secCollectiveCoordinates} describes a variational approach in terms of
collective coordinates, which provides first quantitative answers. 
(iii) How robust are the stable cases against small experimental
imperfections? In Sec.~\ref{secStability}, we develop a linear
stability analysis for periodic perturbations using Floquet theory,
which proves to be in excellent agreement with the numerics.

%----------------------------------------------------
\section{Cyclic-time solutions for wide wave packets}\label{secSymmetry}
In all of the following, we consider wide wave packets, which span many
lattice sites, as obtained by adiabatically loading an extended BEC from a shallow trap into 
an optical lattice \cite{Gustavsson2008}. The first effect of the
mean-field tight-binding equation \eqref{eqTightBinding} is to imprint
a phase factor $\exp(-i F n t)$ onto each amplitude $\Psi_n$.
Such a phase factor can be separated from a smooth envelope
$A(z,t)$ centered on the moving reference point $x(t)$, by the ansatz  
\begin{equation}\label{eqSmoothEnv}
\Psi_n(t) = e^{i p(t) n} A(n-x(t),t) e^{i \phi(t)} \ .
\end{equation} 
The equation of motion obeyed by the envelope $A(z,t)$ is found by
Taylor-expanding the hopping terms of Eq.~\eqref{eqTightBinding} as 
$\Psi_{n \pm 1} = e^{\pm i p}(A \pm \partial_z A +\partial_z^2 A/2) e^{i p n + i
  \phi(t)}$, with third and higher-order derivatives of $A$ assumed to be
negligible (Sec.~\ref{secValiditySmoothEnverlope} below discusses the
validity of this
assumption).
The force term is taken care of by choosing 
\begin{equation} \label{pFt}
p(t) = - Ft \ ,  
\end{equation} 
for the initial condition $p(0)=0$. 
The first spatial derivative $\partial_z A$ can be eliminated by setting
\begin{equation}\label{xt}
x(t) =\frac{2}{F}\, [\cos(Ft)-1]\ , 
\end{equation} 
such that $x(0)=0$, and $\phi(t) = 2\sin(Ft)/F$ with $\phi(0)=0$
without loss of generality. These choices describe the uniform motion of the
quasi-momentum across the Brillouin zone and the resulting
BO in real space. During this oscillation, the envelope
is found to obey the nonlinear
Schr\"{o}dinger equation (NLSE) 
 \begin{equation}
 \label{eqAmplitudeEq}
  i \partial_t A = -\,\frac{1}{2m(t)} \, \partial_z^2 A + g(t) |A|^2 A \ , 
 \end{equation}
with the oscillating inverse mass $m(t)^{-1} = 2\cos(Ft)$. 
Before analyzing the effect of a modulated interaction $g(t)$ in Sec.~\ref{secBlochPeriodicBoundedTime}, we
first describe the usual linear BOs in the absence of interaction
$(g=0)$, while paying particular attention to the internal breathing dynamics.  

%-----------------------------------
\subsection{Linear Bloch oscillation with breathing}
\label{linearBO} 

In the linear case $g=0$, Eq.\ \eqref{eqAmplitudeEq} is the Schr\"odinger equation for a free particle 
with oscillating inverse mass $m(t)^{-1} = 2\cos(Ft)$. 
This problem can be mapped to the even simpler case of constant mass by introducing the cyclic time 
\begin{equation}\label{eta}  
\eta(t) = \frac{\sin Ft}{F}\ . 
\end{equation} 
Since $\partial_t\eta = \cos (Ft) $, the oscillating mass drops out of
the resulting equation of motion 
$  i \partial_\eta \tilde A = - \partial_z^2\tilde A$ for $\tilde A(z,\eta(t)) =
A(z,t)$.  This is the simplest free-particle Schr\"odinger equation,
whose solution reads 
$\tilde A_k(\eta) = \exp(-i k^2 \eta) \tilde A_k(0)$ in momentum representation. 
The real-space solution at cyclic time $\eta$ 
is the initial wave packet $\tilde A(z,0)=A(z,0)$ 
propagated with the unitary evolution operator in position
representation, which is a Gaussian. 
Under this evolution, an initial Gaussian envelope $\tilde A(z,0) = \Psi_z(0)$ such as \eqref{Gaussian} stays Gaussian, 
\begin{equation}
\tilde A(z,\eta) = \frac{\sqrt{\sigma_0}}{\sqrt[4]{2\pi}\tilde\sigma(\eta)}
\exp\left[ -\frac{z^2}{4\tilde\sigma(\eta)^2}\right] \ . 
\end{equation} 
The complex width $\tilde\sigma(\eta)^2 =\sigma_0^2+i\eta$ is monotonic
in the cyclic time $\eta$. But expressed in the physical time $\eta(t)
= \sin(Ft)/F$, the evolution is necessarily periodic, 
\begin{align}\label{eqLinearBO}
A(z,t) &= \frac{\sqrt{\sigma_0}}{\sqrt[4]{2\pi}\hat\sigma(t)}
\exp\left(-\frac{z^2}{4\hat\sigma(t)^2}\right) \ ,
\end{align}
where $\hat\sigma(t)^2 =\sigma_0^2+i\sin(Ft)/F$. This solution describes a
wave packet centered at $z=0$ with variance
\begin{equation} \label{breathing}
\sigma(t)^2 = \int\rmd z\, z^2|A(z,t)|^2 
= \sigma_0^2+ \frac{\sin(Ft)^2}{F^2\sigma_0^2} \ . 
\end{equation} 
The wave packet broadens only initially.
At $Ft = \pi/2$, i.e.\ after the first quarter of the Bloch cycle, the
mass changes sign and the time evolution of the width is reversed.
At the edge of the Brillouin zone $F t=\pi$, the wave packet recovers its original shape.
Thus, the wave packet shows perfectly periodic breathing on top of the
BO; instead of dispersing, it remains localized due to the combination
of lattice and tilt.
The relative amplitude of the breathing \eqref{breathing}
is $1/(F\sigma_0^2)^2$, which should be very small for the
smooth-envelope equation \eqref{eqAmplitudeEq} to be valid. 

The above discussion of the linear BO is based on the NLSE
\eqref{eqAmplitudeEq}, which neglects higher derivatives, i.e., assumes that the
wave packet is smooth and wide. As we will discuss in Sec.\
\ref{secValiditySmoothEnverlope}, the linear BO remains periodic beyond that
assumption, even for very narrow wave packets. With decreasing width, the
breathing increases and the real-space Bloch amplitude is reduced until it
approaches zero \cite{Hartmann2004,Dominguez-Adame2010}.

\subsection{Bloch-periodic interaction}\label{secBlochPeriodicBoundedTime}
As shown already by the noninteracting solution~\eqref{eqLinearBO}, a
Bloch-oscillating wave packet can display rich internal dynamics with  
initial broadening, provided that it recovers its initial state at the
end of the Bloch period.
To begin with, we recall the cyclic-time argument developed in Ref.\ \cite{Diaz2010} and
identify those modulations $g(t)$ which are Bloch periodic and guarantee stable BOs.
Later, in Sec.\ \ref{manymoresol}, we extend this reasoning to arbitrary (rational) frequency ratios.

\subsubsection{General case}\label{secBlochPeriodicBoundedTime_general}
Motivated by the BO stability for $g(t) \propto \cos(F t)$ 
observed in Fig.~\ref{fig1}, we consider the class of Bloch-periodic
interactions that are a product of 
$\cos(F t)$ and any function  $P(\eta)$ of the cyclic time $\eta = {\sin(Ft)}/{F}$ alone:  
\begin{align}
\label{eqStable1}
g(t) = \cos(F t)  P(\eta(t))\ .
\end{align}
Notably, this family includes the higher harmonics $g(t) = g_0 \cos[(2n+1)Ft]$
and $g(t) = g_0 \sin(2n F t)$ for all integer $n$ (as well as all linear
combinations thereof), because 
they can always be brought into the form
\eqref{eqStable1} with the help of trigonometric identities. 
$\tilde A(z,\eta(t)) = A(z,t)$ then obeys the equation of motion 
\begin{align}
\label{eqAmplitudeEqBoundedTime}
 i \partial_\eta \tilde A= & - \partial_z^2 \tilde A + 
 P(\eta) |\tilde A|^2 \tilde A \ ,
\end{align}
which depends only on the cyclic time  $\eta$. And no 
matter the detailed form of its solution $\tilde A(z,\eta)$, 
since $\eta(t)$ is a \emph{periodic function of
time}, the solution $A(z,t)$ must be periodic as well. 
Just as in the linear case discussed in Sec.~\ref{linearBO} 
above, the envelope time evolution
over the first quarter of every Bloch period will be exactly reversed during 
the second quarter.  

The family of interactions~\eqref{eqStable1} includes the cases $g(t) = \pm g_0 \cos(F
t)$, but not $\pm g_0\sin(F t)$, which is a first explanation of the strikingly different behavior
exhibited in Figs.~\ref{fig1} (a) and (b). In the latter case, the
equation of motion cannot be written in terms of $\eta$ alone, so
that the cyclic-time argument does not apply. Of course, this fact by
itself does not necessarily imply that modulation (b) is unstable, but 
Secs.~\ref {secCollectiveCoordinates} and \ref{secStability} below
will show that this is indeed the case. 

\subsubsection{Special case: rigid soliton}
\label{secSoliton}
Let us for a moment consider the NLSE \eqref{eqAmplitudeEq} from a
quasistatic point of view, i.e.\ take mass $m$ and interaction $g$ as constant. 
In the usual case of positive mass, a linear wave packet disperses.
This dispersion has to be compensated by an attractive interaction
in order to obtain a stationary wave packet.
For a negative mass (e.g.\ quasi-momentum close to the band edge), repulsive interaction is needed in order to prevent the
wave packet from contracting. 
If mass and interaction have opposite signs, the NLSE \eqref{eqAmplitudeEq}
admits a stable soliton solution \cite{Konotop2002}:
\begin{equation} 
\label{eqSoliton.mod}
A(z,t) = \frac{1}{\sqrt{2 \xi}}\,\frac{1}{\cosh\left(z/\xi\right)} \,
e^{-i \omega t} \ , 
\end{equation}
with a characteristic width $\xi = - 2/(g m)>0$.

Such a soliton configuration can be maintained during the BO if the
interaction $g$ is modulated such that its sign is always opposite to
the sign of the mass~\cite{Bludov2009}. 
However, the mere existence of a soliton configuration at all times is not sufficient for the preservation of the wave packet.
If the equilibrium width $\xi(t) = -2/[m(t) g(t)]$ changes rapidly in
time, the soliton cannot evolve adiabatically, and internal degrees of
freedom are excited, which will finally destroy the soliton. 
Therefore, the simplest way to preserve a long-living wave packet is to have no
internal dynamics at all. 
To this aim, the interaction parameter
is modulated such that the equilibrium width $\xi_0$ is constant, i.e.,
$g_{\rm r}(t)=-|g_{\rm r}| \cos(Ft)$ where $|g_{\rm r}|=4/\xi_0$. 
Hereafter, this case will be referred to as \emph{rigid soliton}.

In fact, our previous discussion shows that the rigid soliton is but a
special member of the more general family of stable solutions. 
The stable interaction of Fig.~\ref{fig1}~(a), $g(t)=+|g_0|\cos(Ft)$,
enhances the breathing of the linear BO, while the  
modulation $g(t)=-|g_0|\cos(Ft)$  tends to suppress it, even causing antibreathing for 
$|g_0|>|g_{\rm r}|$. 
Clearly, the $-$cos case fulfills the soliton stability criterion $m(t) g(t) < 0$ for all 
times, whereas the $+$cos case does not.  
The preceding time-reversal argument, however, assures that both of them lead to undamped BOs---at least within the approximations underlying the NLSE~\eqref{eqAmplitudeEq}. 
Thus, while the rigid-soliton criterion is sufficient for stability, it is by no means necessary.

%----------------------------------------------------------
\subsection{Bloch-commensurate interaction}
\label{manymoresol}

The class of functions \eqref{eqStable1} covers all stable modulations
$g(t)$ that are higher harmonics of the Bloch frequency, i.e.\
with frequency  $\omega= l F$, $l \in \mathbb{N}$.
Let us generalize the cyclic-time argument to Bloch-commensurate
interactions $g(t)$ evolving at a frequency $\omega= l F/\nu$    
(i.e.\ with period $\nu\TB/l$) where $\nu,l \in \mathbb{N}$ are coprime.
The common period with $\cos(F t)$ is then $T=\nu \TB$. 

We seek to factorize both the oscillating mass and interaction in
the form 
\begin{align}
\cos (Ft)&  = \dot\eta M(\eta)\ , \label{factorm}\\ 
g(t) & = \dot\eta P(\eta)\ , \label{factorg}
\end{align} 
in terms of a suitable cyclic time $\eta(t)$ and otherwise arbitrary
functions $M$ and $P$. Then, the $\dot\eta$ drops out of Eq.\ 
\eqref{eqAmplitudeEq}, which can be written in terms of the cyclic
time $\eta$ only:
\begin{align}
\label{eqAmplitudeEqBoundedTime2}
 i\partial_\eta \tilde A = & - M(\eta) \partial_z^2 \tilde A + 
 P(\eta) |\tilde A|^2 \tilde A\ .
\end{align}
Its solution may be slightly more complicated than that of
Eq.~\eqref{eqAmplitudeEqBoundedTime}, but, again, 
$A(z,t) = \tilde A(z,\eta(t))$ is periodic because of the periodicity of $\eta(t)$.

In order to achieve the factorization \eqref{factorm} and \eqref{factorg},
we observe that trigonometric identities permit writing
$\cos F t =\pm \sin \nu\tau = \pm \sin\tau \, M_{\nu}(\cos\tau)$, with a
polynomial $M_{\nu}$ of degree $\nu-1$, in terms of one of the slower angles
$\tau$ defined via $\nu \tau  = Ft  \pm \pi/2 \mod 2\pi$, 
or equivalently:
\begin{equation}\label{deftau}
\tau = \tau_{\nu j}(t) = \frac{1}{\nu}\Bigl[Ft - \frac{\pi}{2}(2j+1)\Bigr]\ ,\quad j \in \mathbb{Z} \ .
\end{equation}
Defining $\eta = \cos\tau$ and $M(\eta) =  (-1)^j(\nu/F)M_{\nu}(\eta)$ then realizes
Eq.\ \eqref{factorm}. And from Eq.\ \eqref{factorg}, we conclude that all
modulations of the form 
\begin{equation}\label{eqStable2} 
g(t) = \sin\tau \, \tilde P(\cos\tau)\ , 
\end{equation}
with arbitrary $\tilde P(\eta) = -(F/\nu)P(\eta)$, guarantee stable modulated Bloch oscillations. 
This is equivalent to two conditions \cite{Gaul2009}:
\begin{itemize}
\item $g(t)$ has two common zeros with $\cos(F t)$, namely
  $t=t_0,t_0+T/2$, the
  zeros  of $\sin \tau$ [see Eq.~\eqref{deftau}].
\item $g(t)$ is odd with respect to these points, $g(t_0 + t') = -g(t_0 - t')$, and similar for $t_0+T/2$.
\end{itemize}

%-------------------------------------------------------------------------
\begin{figure}%[bt]
\includegraphics[angle=-90,width=\linewidth]{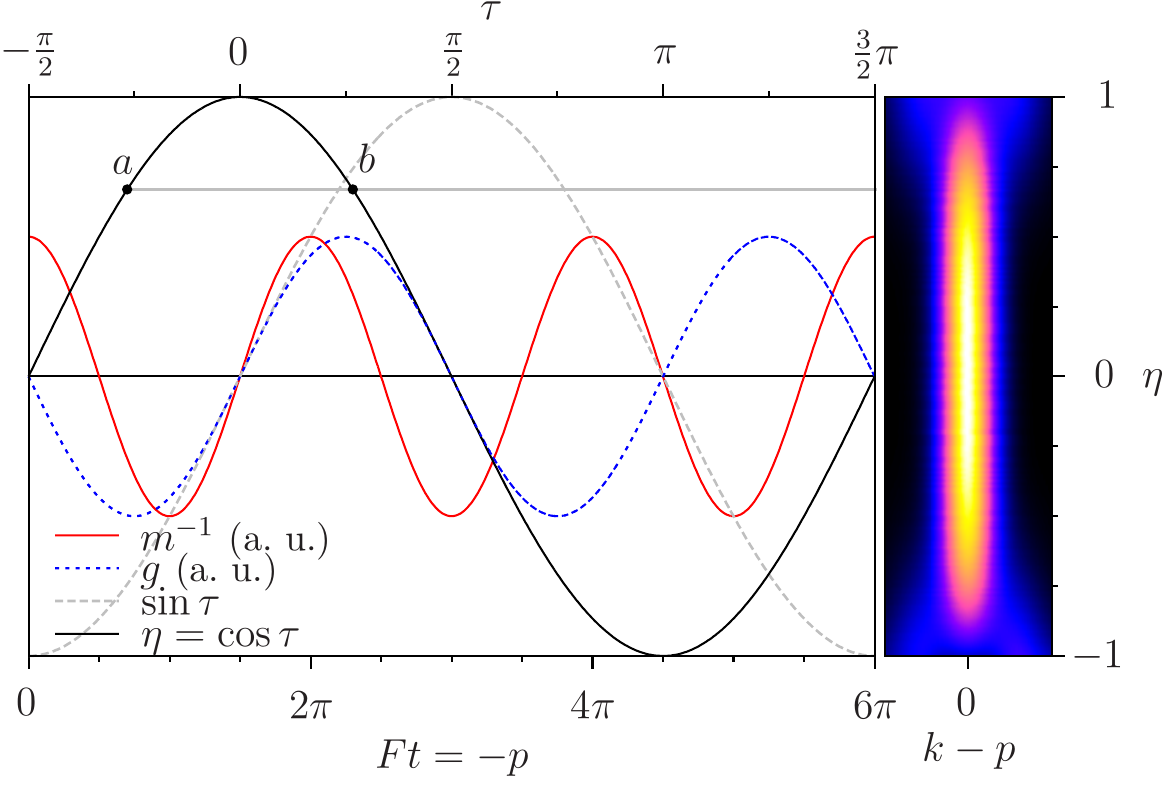}
\caption{
(Color online) Time evolution scheme for a stable modulation $g(t)$ with frequency ratio $l/\nu = 2/3$.
The common nodes of $m(t)^{-1} = 2 \cos(F t)$ and $g(t)$ are shared by $\sin \tau$,
with $\tau = \tau_{3 1}(t)=(Ft-3\pi/2)/3$ [cf.\ Eq.\ \eqref{deftau}].
As required by Eq.~\eqref{eqStable2}, the interaction can be written as
$g(t) = 2g_0 \sin\tau \cos \tau = g_0 \sin\bigl(2 F (t - 3\TB/4)/3\bigr)$.
Thus, the envelope function is a function of $\eta = \cos \tau$ only and is consequently periodic in $t$.
The right panel shows the momentum space density $|\Psi_{k-p}|^2$, obtained by numerical integration of
Eq.~\eqref{eqTightBinding} with initial condition \eqref{Gaussian} and $\sigma_0 = 10$.
The points in time with the same value of $\eta$, like $a$ and $b$, show the same distribution.
Parameters are $g_0=5$, $F=0.2$.}
\label{figBreathingk}
\end{figure}
%-------------------------------------------------------------------------------

Let us illustrate how to construct such a stable modulation via the
example of a  $g(t)$ that is harmonically modulated with frequency
ratio $\omega/F = l/\nu= 2/3$, as shown in Fig.~\ref{figBreathingk}.
The commensurability $\nu = 3$ and the choice $j=1$
determine $\tau = \tau_{3 1}(t) = (Ft-3\pi/2)/3$, %[Eq.\ \eqref{deftau}], 
and from there the cyclic time $\eta  = \cos\tau$. 
According to Eq.\ \eqref{eqStable2}, 
a harmonic modulation with the desired frequency ratio is $g(t) = g_0 \sin 2\tau = 2 g_0 \sin \tau  \cos \tau$. 
Then, the wave function $A$ is a function of $\eta$ only, as shown in
the right panel of Fig.~\ref{figBreathingk}.

Similarly, any harmonic modulation that is commensurate with the Bloch frequency, $\cos[(l/\nu) F t + \delta]$, can be constructed.
If the interaction writes
\begin{align}\label{stableg.eq}
g(t) = g_0 \sin\bigl(l \, \tau_{\nu j}(t)\bigr) =
g_0 \sin\biggl\{ 
\frac{l}{\nu} 
 \left[Ft - \frac{\pi}{2}(2j+1)\right]
\biggr\} 
\end{align}
(with $\nu, l ,j \in \mathbb{N}$), then the BO is stable. In Ref.~\cite{Gaul2009}, this result has been derived in a different but equivalent way.
The class of functions
\eqref{stableg.eq} covers all frequencies commensurate with the Bloch
frequency,  if only the phase with respect to the BO is adjusted
correctly (see Sec.~\ref{secRobustnessCommensurtate} and
Fig.~\ref{figMap} below).

%----------------------------------------------------------
\subsection{Beyond the smooth-envelope approximation}\label{secValiditySmoothEnverlope}
The cyclic-time argument discussed above is based on the NLSE \eqref{eqAmplitudeEq}, 
where higher-than-second-order spatial derivatives have been neglected.
We still need to estimate the effects caused by the third-order
derivative.

In the \emph{linear case},  each Fourier component can be treated
separately, just as in Sec.\ \ref{linearBO}. The  
third derivative shows up in 
\begin{eqnarray}
\label{eq3rdDerivative}
i \partial_t A_k &=& k^2 \left( \cos F t + \frac{k \sin F t}{3} \right) A_k 
\nonumber \\
&=& k^2 \sqrt{1+k^2/9} \, \cos(F t - \phi_k) A_k \ , 
\end{eqnarray}
with $\tan\phi_k = k/3$, which can be easily solved 
in terms of its own cyclic time 
$\eta_k(t) = \sin(F t-\phi_k)$.
The result is, of course, periodic, 
\begin{equation}\label{eqA0}
 A_k^{(0)}(t) = A_k(0)\,e^{-i\varphi_k(t)}\ ,
\end{equation}
where
\begin{equation}
\varphi_k(t)=\frac{1}{F}\left[k^2\sin F t + \frac{k^3}{3}(1-\cos F t)\right]\ .
\end{equation}
The initial Gaussian wave packet \eqref{Gaussian} with $A_k(0) =
\sqrt[4]{2/\pi}\sqrt{\sigma_0}\exp(-k^2\sigma_0^2)$ 
restricts the relevant values of $k$ to be of the order $\sigma_0^{-1}$.
Thus, corrections due to the third derivative are small for wide wave packets.
The superposition of all $A_k(t)$ leads to a reduction of the bare real-space Bloch amplitude \eqref{xt}
by $\langle z \rangle \approx - [\cos (F t) - 1]/(F\sigma_0^2)$,
which is indeed the leading order of the full result given in Ref.\ \cite{Hartmann2004}.

\paragraph*{Interaction $g(t)=g_0 \cos (F t)$.}
What happens if interactions, as in Sec.\ \ref{secBlochPeriodicBoundedTime}, are included into Eq.\ \eqref{eq3rdDerivative}?
The interaction term 
$g(t)\int {\rm d}k' {\rm d}q A_{k-q}A_{q-k'}^*A_{k'}/2\pi$
mixes all components, which consequently cannot be solved separately.
On the other hand, also the global cyclic-time argument from Sec.\
\ref{secBlochPeriodicBoundedTime} is broken by the factor $(k/3)\sin(F
t)$ from the third derivative. 
Thus, we expect a decay of the BO, 
even if the interaction satisfies the cyclic-time criterion \eqref{eqStable1}.
This decay should scale as the product of $g$ and some wide-wave-packet parameter related to $\sigma_0^{-1}$.

We estimate this decay analytically by considering the first-order correction
$A_k^{(1)}(t) = a_k(t) \exp[-{i}\varphi_k(t)] $ to Eq.~\eqref{eqA0}, due to $g(t) = g_0 \cos(F t)$.
To this end we integrate the equation of motion
\begin{equation}
 \dot a_k = -i e^{i \varphi_k(t)} g_0 \cos(F t) \int \frac{{\rm d}k' {\rm d}q}{2\pi}
       A^{(0)}_{k-q} \bigl(A^{(0)}_{q-k'}\bigr)^*A^{(0)}_{k'}\ .
\end{equation}
We take advantage of the wide wave packet and expand systematically in the parameter 
$\sigma_0^{-1} \sim k$.
The interaction integral evaluates to $(2\pi)^{-1/4}(3\pi\sigma_0)^{-1/2} e^{-k^2\sigma_0^2/3}$ 
plus terms of second order in $k$ and $\sigma_0^{-1}$. 
The time integral of $e^{i \varphi_k(t)} \cos(F t)$ over the Bloch period is nonzero only due to 
the phase shift $\phi_k \approx k/3$ between $\varphi_k(t)$ 
and $\cos (F t)$.
It is expressed in terms of the Bessel function of the first kind  $J_1(k^2/F) \approx k^2/2F$.
Finally we get
\begin{align}\label{eqA1}
 a_k(\TB) %-a_k(0) 
=  \left(\frac{\pi}{2}\right)^{\frac{1}{4}} 
\frac{-g_0 k^3}{3\sqrt{3\sigma_0}F^2 }
\exp\Bigl(-\frac{k^2\sigma_0^2}{3}\Bigr)\ .
\end{align}
The relevant values of $k$ are cut off by the exponential and scale as $k\propto \sigma_0^{-1}$.
No matter what the sign of $g_0$, the net growth of the first-order correction \eqref{eqA1} after one Bloch period deforms the wave packet and destroys the periodic dynamics of the BO.
This leaves us with the general scaling of the life time
\begin{equation}\label{eqLifetimeSacling}
\frac{1}{T_{\rm life}} \sim \frac{|a_k(\TB)|}{\TB} \propto \frac{|g_0|}{F \sigma_{0}^{3.5}}\ ,
\end{equation}
assuring very long life times for sufficiently wide wave packets.

This result proves to be quite reliable, as we have checked by means of a direct integration of the tight-binding equation of motion \eqref{eqTightBinding} for several sets of parameters.
Numerically we defined the life time as the time when the momentum variance has doubled with respect to its initial value.
This includes averaging over the contributions of many modes and dynamics that go already quite far away from the original perturbative perspective.
Still, we find this life time to scale as predicted by Eq.~\eqref{eqLifetimeSacling}, with proportionality factor
$\approx0.2$. 
% around our standard parameters $F=0.2$, $g(t)=g_0 \cos(F t)$, $\sigma_0=10$.

%----------------------------------------------------
\section{Collective coordinates}\label{secCollectiveCoordinates}
From Sec.~\ref{secSymmetry}, we know  which modulations $g(t)$ should lead to stable BOs.
But we wish to gain more quantitative information on how position and momentum distributions depend on
the modulation. Moreover, we would like to describe the time evolution
also in the unstable cases. Toward this aim, we employ the
CC approach, as introduced in Ref.\ \cite{Trombettoni2001}. 

%----------------------------------
\subsection{Equations of motion}

The equation of motion (\ref{eqTightBinding}) derives as $i\dot\Psi_n = \partial H/\partial\Psi_n^\ast$  
from the mean-field Hamiltonian  
\begin{equation}\label{Hamiltonian.eq}
H = \sum_n \left[-(\Psi_{n+1}\Psi_n^\ast+c.c.) + F n |\Psi_n|^2 + 
\frac{g(t)}{2}|\Psi_n|^4 \right],  
\end{equation}
where $\Psi_n$ and $i\Psi_n^*$ are canonically conjugate
variables. Instead of describing all these amplitudes, we restrict the
number of degrees of freedom and parametrize the dynamics of a smooth
wave packet by its centroid $x(t) = \mv{n}= \sum_n n
|\Psi_n(t)|^2$ and variance $w(t) =  \mv{[n-x(t)]^2}$. One also needs their
respective conjugate momenta $p(t)$ and $b(t)$, defined by their
generating role  $-i\partial_p\Psi_n= n \Psi_n$ and similarly
for $b$. Thus, we employ the ansatz 
\begin{align}\label{psiA.eq}
\Psi_n(t) = \frac{1}{\sqrt[4]{w}}\, {\mathcal{A}} 
\left(\frac{n-x}{\sqrt{w}}\right) 
e^{i p n +i b (n-x)^2
} \ ,
\end{align}
with an even envelope function 
${\mathcal{A}}(u)$ that is normalized according to
$\int {\rm d}u |{\mathcal{A}}(u)|^2 = 1$ and 
$\int {\rm d}u \, u^2 |{\mathcal{A}}(u)|^2=1$.
The assumptions underlying the CC description \eqref{psiA.eq}  differ slightly 
from the smooth-envelope ansatz, Eq.~\eqref{eqSmoothEnv}. Keeping a
fixed wave-packet shape  $\mathcal A(u)$ is of course more restrictive.
On the other hand, the centroid $x$ is now a free dynamical variable,
which gives enhanced 
flexibility compared to the purely kinematic $x(t)$ of Eq.\ \eqref{xt}. 

Inserting the CC ansatz \eqref{psiA.eq} into the
Hamiltonian \eqref{Hamiltonian.eq}, 
Taylor-expanding the discrete gradient to second order and performing
the sum as a continuous integral, one finds the effective Hamiltonian
\begin{equation}\label{Heff.eq}
\Heff  = F x - 2 \cos p \left(1- \frac{K + 4 b^2w^2}{2w}\right)  + 
I \, \frac{g(t)}{\sqrt{w}}\ ,
\end{equation}
with the kinetic integral $K = \int {\rm d}u |\mathcal{A}'(u)|^2$ 
and the interaction integral $I = (1/2) \int {\rm d}u |\mathcal{A}(u)|^4$.
In Table~\ref{tabBO_cc}, these are given for a Gaussian and for a soliton-shaped wave packet.

%-------------
\begin{table}[b]
\caption{Collective-coordinates parameters for Gaussian and soliton wave packets\label{tabBO_cc}}
\centerline{
\newcommand\Tp{\rule{0pt}{3.6ex}}
\newcommand\Bt{\rule[-1.2ex]{0pt}{0pt}}
 \begin{tabular}{l|ccc}
  %shape 
\Bt	& ${\mathcal{A}}(u)$		& $K$		& $I$ \\
 \hline  %---------------------------------------------------------------
  Gaussian \Tp	& $(2\pi)^{-\frac{1}{4}} \, \displaystyle{e^{-\frac{u^2}{4}}}$
				& $\frac{1}{4}$	& $\frac{1}{4\sqrt{\pi}}$	\\
  soliton	& $\sqrt{\frac{\pi}{4\sqrt{3}}} \, {\left[{\cosh\left(\frac{\pi \, u}{2\sqrt{3}}\right)}\right]^{-1}}$
				& $\frac{1}{4}\left(\frac{\pi}{3}\right)^2$
						& $\frac{1}{4\sqrt{\pi}} \left(\frac{\pi}{3}\right)^\frac{3}{2}$
 \end{tabular}
}
\end{table}
%-----------

By construction, the CC variables obey the canonical equations of motion 
\begin{align}
\dot p &=- \frac{\partial \Heff}{\partial x} = -F\ ,    \label{pdot} \\
\dot x &= \frac{\partial \Heff}{\partial p} = 2 \sin p \left[1-\frac{K+4b^2w^2}{2w}\right] \ , \label{xdot}\\
\dot b &= - \frac{\partial \Heff}{\partial w} = \frac{K-4w^2b^2}{w^2}\cos p +  \frac{I\,  g(t)}{2 w^{3/2}}\ , \label{bdot} \\    
\dot w &= \frac{\partial \Heff}{\partial b} = 8 w b \cos p\ .   \label{wdot}  
\end{align}\label{eqsCC}%
In our study, the following initial conditions are considered: $x(0)= 0$, $p(0)=0$, $w(0) = \sigma_0^2$, $b(0)=0$.
Equation\ \eqref{pdot} shows that the driving term in all the equations is
$p(t) = -Ft$, as already used in Eq.~\eqref{pFt}, and this independently of the interaction. In other words, the
Bloch period is not affected by the atom-atom interaction. 
Furthermore, the dynamics of $b(t)$ and $w(t)$ is completely defined
by the autonomous Eqs.\ \eqref{bdot} and \eqref{wdot}, whose solution 
then determines the centroid motion according to Eq.~\eqref{xdot}.

In cold-atom experiments, time-of-flight images provide the momentum
distribution of the atomic cloud. Therefore, it is appropriate to
study not only the average (quasi-)momentum $p$, but also the momentum
variance,  
whose growth in time serves as a good indicator for the decay of the wave packet.
Actually, the momentum variance appears naturally already in the effective Hamiltonian \eqref{Heff.eq}.
Indeed, the kinetic energy contribution is nothing but the mean-field expectation
$-\mv{e^{i\hat p} + e^{-i\hat p}}$ of (minus twice the real part of) the discrete translation operator.
For a sufficiently narrow momentum distribution this contributes  
\begin{eqnarray} 
\mv{\cos\hat p} &\approx &  \cos p \left(1-\frac{(\Delta p)^2}{2}\right)\ .
\end{eqnarray} 
Comparing with the second term on the right hand side of Eq.\ \eqref{Heff.eq}, we recognize $(\Delta p)^2 = K/w + 4 b^2 w$. 
In passing, we note that this implies that $K=(\Delta x)^2 (\Delta p)^2  - 4 b^2 (\Delta x)^4$ is a constant of motion.
Since the wave packet has a fixed shape, this quantity can only be the surface of the uncertainty ellipse, 
$K=(\Delta x)^2 (\Delta p)^2  - (\Delta x p)^2$, so that we can identify $4b^2= (\Delta x p)^2/(\Delta x)^4$.

In terms of the momentum variance  $r:=(\Delta p)^2$, 
the equation of motion \eqref{xdot} for the centroid looks slightly simpler:
\begin{align}
\dot x & =  - 2\left(1-\frac{r}{2}\right) \sin (Ft)  \ .
\label{dotx2}
\end{align}
The momentum variance obeys the equation of motion 
\begin{align}
\dot r & = 4 I g(t) b w^{-1/2} \label{dotr2}\ . 
\end{align}

In the linear case $g=0$ the momentum variance is a constant of motion, $(\Delta p)^2=r_0 =
K/\sigma_0^2$.
\footnote{The entire momentum distribution
  remains unchanged, up to the uniform translation $p= -Ft$ across the
  Brillouin zone,
  as can be seen from the reasoning in Sec.\ \ref{secValiditySmoothEnverlope}:
  even if arbitraryly high derivatives are included in the first line of 
Eq.\ \eqref{eq3rdDerivative}, the Fourier components $A_k$ do not mix; each of 
them acquires only a phase factor and $|\Psi_{k-p}|^2 = |A_k|^2$ is stationary.}
Then, Eq.~\eqref{dotx2} integrates to $x(t) = (2/F)\left(1-r_0/2\right) [\cos(Ft) -1]$.
Compared to the lowest-order result, Eq.\ \eqref{xt}, the amplitude of the BO
is found to be reduced by the finite momentum width, in agreement with the exact
solution for Gaussian wave packets of arbitrary width, where the
amplitude is reduced by a factor $e^{-r_0/2}$ \cite{Hartmann2004}.
Furthermore, Eqs.\ \eqref{bdot} and \eqref{wdot} yield the exact solution 
$w(t) = \sigma_0^2 + 4 K \sin^2(Ft)/(F \sigma_0)^2$, which agrees with Eq.~\eqref{breathing}. 

\subsection{Interaction effects}\label{sSecCCInteraction}
Let us now study the CC equations \eqref{pdot}--\eqref{wdot} in the interacting
case for several different modulations of the interaction parameter $g(t)$.
In all of the following examples, we compare the CC results to the full integration of the tight-binding equation.
Furthermore, we analytically isolate the leading-order effects caused by the interaction.
We take advantage of the wide-wave-packet condition, $F\sigma_0^2\gg 1$,
and treat the interaction $g$ perturbatively.
To zeroth order in $g$ and $1/F\sigma_0^{2}$, we have $w(t) =
\sigma_0^2$ and $b(t)=0$. Then, we compute the leading corrections via
Eq.\ \eqref{bdot} and subsequently Eqs.\ \eqref{dotx2} and
\eqref{dotr2}, or Eq.\ \eqref{wdot}.

\subsubsection{Constant interaction}\label{ssSecConstInteraction}

\begin{figure}%[btp]
\includegraphics[width=\linewidth]{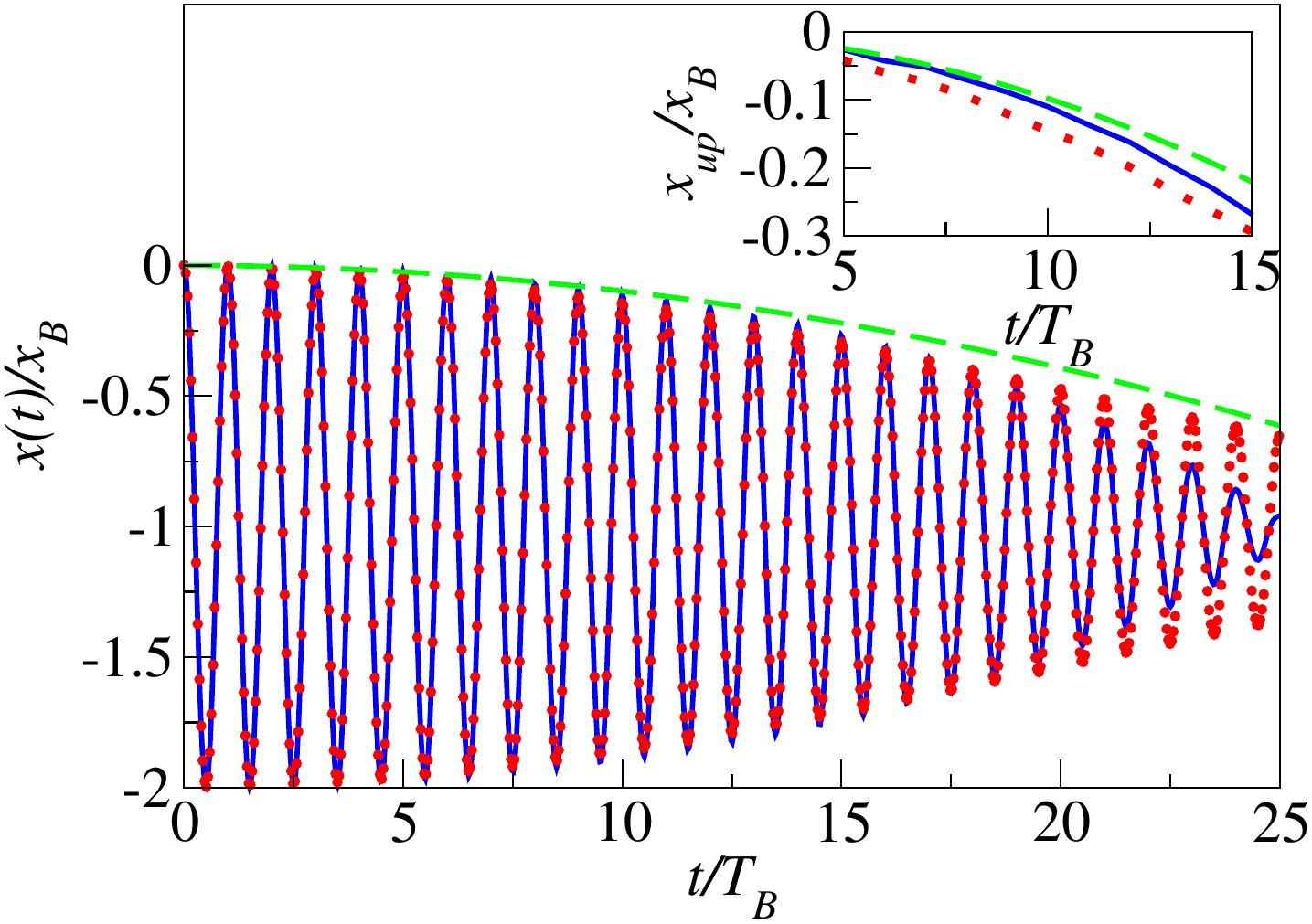}
\caption{(Color online) Centroid motion of a Gaussian wave packet for constant interaction $g(t)=g_0$, obtained 
by numerical integration of the full tight-binding equation of motion~\eqref{eqTightBinding} (red dots) 
and CC approach~(\protect{\ref{pdot}})--(\protect{\ref{wdot}}) (solid blue).
The dashed (green) line shows the upper turning points of the centroid within the perturbative result \eqref{x_g0}, $x_{\rm up} =-(\tavg{r}-r_0)/F$. The inset shows these 
upper turning points on a larger scale.
Parameters are $F=0.2$, $g_0=1.0$ and $\sigma_0=10$. 
\label{figDamping}}
\end{figure}

Constant interaction $g(t)=g_0$ is known to cause momentum broadening
and damping \cite{Trombettoni2001,Witthaut2005}. 
From Eq.\ \eqref{bdot}, we find the averaged linear increase $\tavg{b(t)} = I g_0 t
/(2\sigma_0^3)$. 
Here and in the following, the overline denotes coarse graining over one Bloch period.
Via Eqs.\ \eqref{dotr2} and \eqref{dotx2}, we find 
\begin{align}
\tavg{r(t)} &\approx r_0+ \frac{I^2 g_0^2 t^2}{\sigma_0^4}\label{r_g0} \ , \\
x(t) &\approx \frac{2}{F}
 \left\lbrace
  \Bigl[1-\frac{\tavg{r(t)}}{2}\Bigr]\cos(Ft) - \Bigl[1-\frac{r_0}{2}\Bigr]
 \right\rbrace , \label{x_g0}
\end{align}
which is consistent with Refs.~\cite{Trombettoni2001,Witthaut2005}.
In Fig.~\ref{figDamping}, we compare this perturbative result to the
full CC prediction \eqref{pdot}--\eqref{wdot} and to the results
from the integration of the discrete Gross-Pitaevskii equation
\eqref{eqTightBinding}. 
The approximation \eqref{x_g0} initially agrees nicely with the CC 
prediction~\eqref{pdot}--\eqref{wdot}.
The zoomed envelope of the centroid motion, shown in the inset, reveals,
however, that the CC approach initially ($t\lesssim 15
T_{\rm B}$) underestimates the damping with respect to the full
result. 
This is because the CC ansatz misses momentum broadening and energy losses due to degrees of freedom not included in the ansatz.
At later times $t\gtrsim 20\TB$, the CC approach overestimates the damping. At this time, the CC ansatz already begins to break down, because the wave packet loses its shape [Fig. \ref{fig1} (b)].

%---------------------
%---------------------

%-----------------
\subsubsection{Harmonic modulation}
Let us now use the CC ansatz to understand the examples of the harmonic modulations of the interaction parameter presented in Fig.\ \ref{fig1}.
 
\begin{figure}[tb]
\includegraphics[angle=270,width=\linewidth]{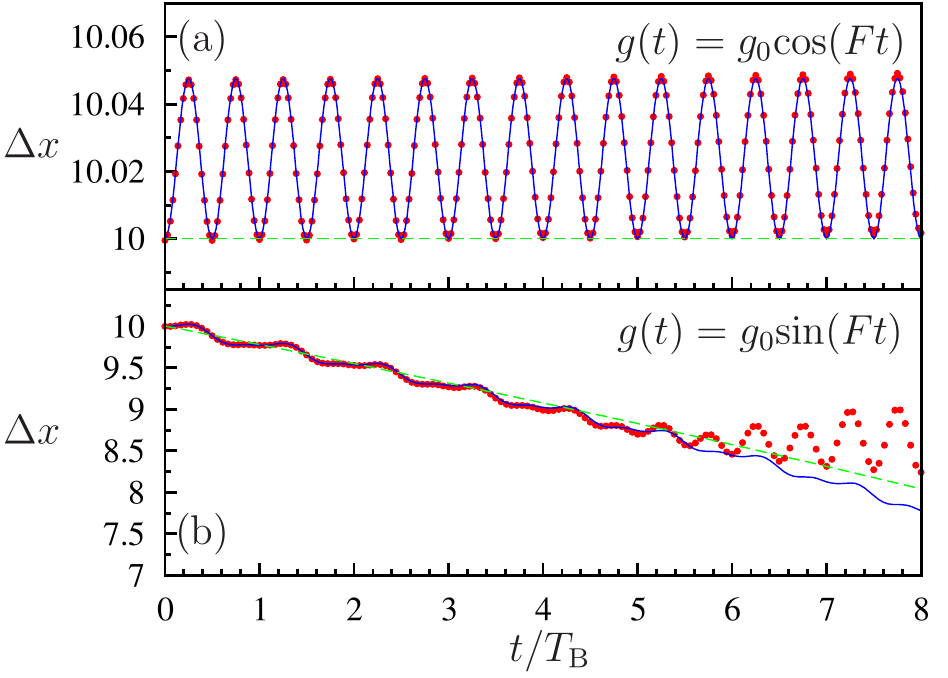}
\caption{(Color online) Width of the Bloch-oscillating wave packets of Fig.\ \ref{fig1} 
($F=0.2$, $\sigma_0=10$, $g_0=1$). %(a) (top) and Fig.\ \ref{fig1}(b) (bottom).
The respective curves show:
$\Delta x$ obtained from the tight-binding equation of motion \eqref{eqTightBinding1} (red dots),
the CC result $\Delta x = \sqrt{w}$ (solid blue),
the estimate $\sqrt{\tavg{ w } }$ from Eq.\ \eqref{eqEstimateContraction} (dashed green).
}\label{figCCwidth}
\end{figure}

\paragraph{Cosine modulation} $g(t) = g_{\rm 0}\cos(F
t)$ [Fig.\ \ref{fig1}~(a)]: BOs are stable, in agreement with the 
cyclic-time argument of Sec.\ \ref{secSymmetry}. 
The breathing of the linear BO is due to the first term in Eq.\ \eqref{bdot}.
The cosine modulation of $g(t)$ in the second term enhances or suppresses this breathing.
Indeed, the breathing amplitude in Eq.\ \eqref{breathing} gets multiplied by a factor $(1 + I\sigma_0 g_{\rm 0} / 2K)$, which in our example of Fig.~\ref{fig1}~(a) 
evaluates to 2.82. Thus, the breathing induced by $g_{0}$ is considerably stronger than the linear breathing.
Nevertheless, both the integration of the CC equations of motion and the analytical result are in perfect agreement with the full tight-binding equation, as shown in the upper panel of Fig.\ \ref{figCCwidth}.
Thus, in this case, where the shape of the wave packet is preserved, the CC approach proves to be a very powerful tool.

\paragraph{Sine modulation} $g(t) = g_{0} \sin(F t)$
[Fig.~\ref{fig1}~(b)]: In the long run, the wave packet looses its
shape, which cannot be accurately described by CCs.
During the first few Bloch periods, however, the wave packet is still
intact and we may use CCs to describe, for example,
the width of the wave packet.  
The interaction $g(t)$ enters via Eq.\ \eqref{bdot} and induces
$b(t) \approx  K \sin(F t) /F\sigma_0^4 - I g_{0} \cos(F t) / 2F\sigma_{0}^{3}$.
Only the second term, proportional to $g_{0}$, has a non-vanishing
time average after multiplying by 
$\cos p =\cos(F t)$ in the equation of motion \eqref{wdot}, which leads to
\begin{align}
\tavg{w(t)} \approx  
\sigma_0^2 - 2 \frac{I g_{0}}{F \sigma_0} \, t \ .\label{eqEstimateContraction}
\end{align}
This prediction is shown in the lower panel of Fig.~\ref{figCCwidth}, together with the full solution of Eqs.\ \eqref{bdot} and \eqref{wdot} and the width extracted from the integration of the tight-binding equation of 
motion~\eqref{eqTightBinding} [shown in Fig.\ \ref{fig1}~(b)].
The CC description fares very well up to $t\approx 5 T_{\rm B}$. At
this time the smooth shape of the wave packet is lost, and deviations from the CC prediction occur without surprise. 

%--------------------------------------------------------------------------

The change of the real-space variance \eqref{eqEstimateContraction}
directly reflects in the change of the momentum variance, $r-r_0 = (\partial_w r) \Delta
w \approx {2 K I g_{0} t}/(F \sigma_0^5)$, driving the decay of the
BO. Note that for a modulation $g(t)\propto -{ \sin F t}$, the momentum variance initially decreases.
On the long run, however, the BO still decays, because the wave packet
does not maintain its shape, just as in the case of $+{\sin F t}$. 

\subsection{Range of validity}
In the above examples, we have demonstrated that, under some
limitations, the CC approach is capable of describing the principal
degrees of freedom, position and variance of a Bloch-oscillating wave packet. 
Since the CC ansatz relies on the shape $\mathcal{A}(u)$ of the wave packet 
[Eq.\ \eqref{psiA.eq}] to be conserved, only the %long-time behavior of
stable situations can be well described on the long run [see Fig.\ \ref{figCCwidth} (a)].
In the unstable cases, CCs can describe only the
initial dynamics, like the contraction of the wave packet shown in Fig.\ \ref{figCCwidth}~(b), but not the decay of its shape. 
For that task, we will pursue a different strategy in the following Section.

%----------------------------------------------------
\section{Linear stability analysis}\label{secStability}
We have already seen [Figs.~\ref{fig1} (b) and \ref{figCCwidth} (b)] that in some
cases perturbations on a length scale much shorter than the width of the wave
packet occur, which ultimately destroys BOs.  The periodically
time-dependent mass and interaction, as appearing in Eq.~\eqref{eqAmplitudeEq},
provide the source of energy for the growth of such perturbations, a phenomenon
known as \emph{dynamical instability} \cite{Fallani2004,Wu2003}.  In the
following, we will employ Floquet theory to detect and to quantify the dynamical
instability.

\subsection{Lyapunov exponents for excitations}
In order to quantitatively describe the growth of short-scale perturbations, we assume that the homogeneous background 
$\Psi_n^{(0)} = \sqrt{n_0}$ is dressed by small fluctuations
$\delta\Phi_n \in \mathbb{C}$.  We insert  
\begin{align}
\Psi_n = \left[\sqrt{n_0} + \delta\Phi_n \right] e^{i p(t) n}
e^{-i\varphi(t)} 
\end{align}
into Eq.~\eqref{eqTightBinding} and expand in 
powers of $\delta \Phi_n$. To zeroth order, $p(t) = -F t$ and
$\dot\varphi = -2 \cos Ft + n_0 g(t)$. This solution describes an
oscillating superfluid flow in a 
spatially homogeneous condensate, i.e.\ a $\delta$-peak in momentum
space that performs perfect BOs, and this no matter how strong the
interaction or how it is modulated. BOs only get dephased by a
combination of inhomogeneity and interaction, as we find by looking at
the first-order equation of motion for $\delta\Phi_n$.
Similarly to the procedure leading to Eq.~\eqref{eqAmplitudeEq}, we
describe $\delta\Phi_{x(t)+z}$ in  a moving reference frame $x(t) = (2/F)[\cos(F t)-1]$.
This eliminates the first derivative of the Taylor expansion of $\delta\Psi_{n \pm 1}$.
Third and higher derivatives are assumed to be small and neglected.
In the present approximation of a homogeneous background, the
equations of motion for different fluctuation Fourier components decouple,
and their real and imaginary parts $\Phi_k = s_k + i\, d_k$ 
then have the coupled equations of motion
\begin{align}
\dot s_k &=  k^2 \cos(Ft)\, d_k\ , \label{dotsk}\\
\dot d_k &= -\left[k^2 \cos(Ft) +2 n_0 g(t)\right]  s_k\ .\label{dotdk}
\end{align}
All these components contribute  to the momentum variance
\begin{equation}
 (\Delta p)^2 = \frac{1}{N n_0} \sum_{k} k^2(|s_k|^2+|d_k|^2)\ ,
\end{equation}
where $N$ is the system size entering the discrete Fourier
transformation. In unstable cases, some fluctuation
with a certain $k$-vector
will possess the largest growth rate and therefore dominate the decay
of BOs. We seek to determine the most unstable mode and its growth rate, or
Lyapunov exponent, in the following. 

In the quasistatic picture, where the mass $\cos Ft = 1/2m_0$ and
interaction $g_0$ are considered constant, the system of coupled
equations \eqref{dotsk} and \eqref{dotdk} can be solved by a
Bogoliubov transformation \cite{Gaul2008,Giorgini1994}: the elementary excitation 
\begin{equation}\label{eqBgTrafo}
c_k = \sqrt{\EBg/\Efr}\,  s_k + i \sqrt{\Efr/\EBg} \, d_k \ ,
\end{equation} 
defined in terms of single-particle dispersion $\Efr=k^2/(2m_0)$ and the Bogoliubov frequency
\begin{equation}\label{eqBgDispersion}
\EBg = \sqrt{{\Efr} ({\Efr} + 2 g_0 n_0)} \ ,
\end{equation} 
has the simple time evolution $c_k(t) \propto \exp({-i \EBg t})$.
The criterion for BO stability then is the following: 
If $m_0$ and $g_0$ are of the same sign, then $\omega_k$ is real for all $k$, and the extended condensate is stable.
If, however, $m_0$ and $g_0$ have opposite signs, then imaginary
frequencies occur for $k< k_* = 2\sqrt{ n_0 |g_0 m_0|}$, indicating
a modulational instability of the extended condensate.  
These modulations lead to the formation of bright solitons \cite{Konotop2002}.
Consistent with this picture is that the critical wave number $k_*$, estimated from the central density $n_0=1/2\xi$, relates to the soliton width: $k_* = 2/\xi$.
But we have already seen in Sec.~\ref{secSoliton} that such a quasistatic
stability criterion does not describe adequately the dynamical stability of
Bloch oscillations with modulated interaction $g(t)$. 

Let us then solve the time-dependent Eqs.~\eqref{dotsk} and \eqref{dotdk}  with
harmonic $g(t)$. These are linear equations with time-periodic coefficients, which makes them accessible for Floquet theory \cite{Markley2004},
provided the frequency of the external modulation $g(t)$ is commensurate with the Bloch frequency $\omega_B=F$. 
Because the driving is periodic, integrating the equations of
motion over a single period $T$  yields all information necessary for the
time evolution over $n \in \mathbb{N}$ periods: 
\begin{equation}\label{sdM}
 \cvect{s_k(t+n T)}{d_k(t+n T)} = M_k^n \cvect{s_k(t)}{d_k(t)}. 
\end{equation}
The monodromy matrix 
\begin{equation} 
M_k = \matr{s_k^1(T)}{s_k^2(T)}{d_k^1(T)}{d_k^2(T)} 
\end{equation}
contains the solution at time $T$ 
starting from the two linearly independent initial
conditions $\{s_k^1(0)=1,d_k^1(0)=0\}$ and
$\{s_k^2(0)=0,d_k^2(0)=1\}$. 
From Eq.~\eqref{sdM}, it is clear that the eigenvalues $\rho_k^\pm$ of
$M_k$ determine the growth of the perturbations.
With the help of Liouville's formula $\det (M_k) = 1$, one finds 
\begin{equation}\label{eqDelta}
\rho_k^\pm = (\mathrm{tr}M_k/2) \pm \sqrt{(\mathrm{tr} M_k/2)^{2} - 1}\ .
\end{equation}
The Lyapunov exponent $\lambda_k =T^{-1} \ln\left[\max( | \rho_k^{+} | , | \rho_k^{-} | )\right]$ then characterizes the exponential growth of the amplitudes $s_k,d_k \sim e^{\lambda_k t}$.
In the following, we explore the consequences of this description in some particularly relevant
cases. 

\subsection{Robustness with respect to perturbations} 
\label{secRobustnessCommensurtate}

\subsubsection{Stability map} 
BOs are stable if the Lyapunov exponent $\lambda_k$ vanishes for all $k$ \cite{Gaul2009}.
This is indeed the case if $g(t)$ fulfills Eq.~\eqref{eqStable2},
because the cyclic-time 
argument from Sec.~\ref{secSymmetry} applies also in the present scope:
All fluctuations $s_k$ and $d_k$ evolve periodically in time, and thus
do not grow exponentially. 

\begin{figure}[tb]
\includegraphics[width=\linewidth,clip]{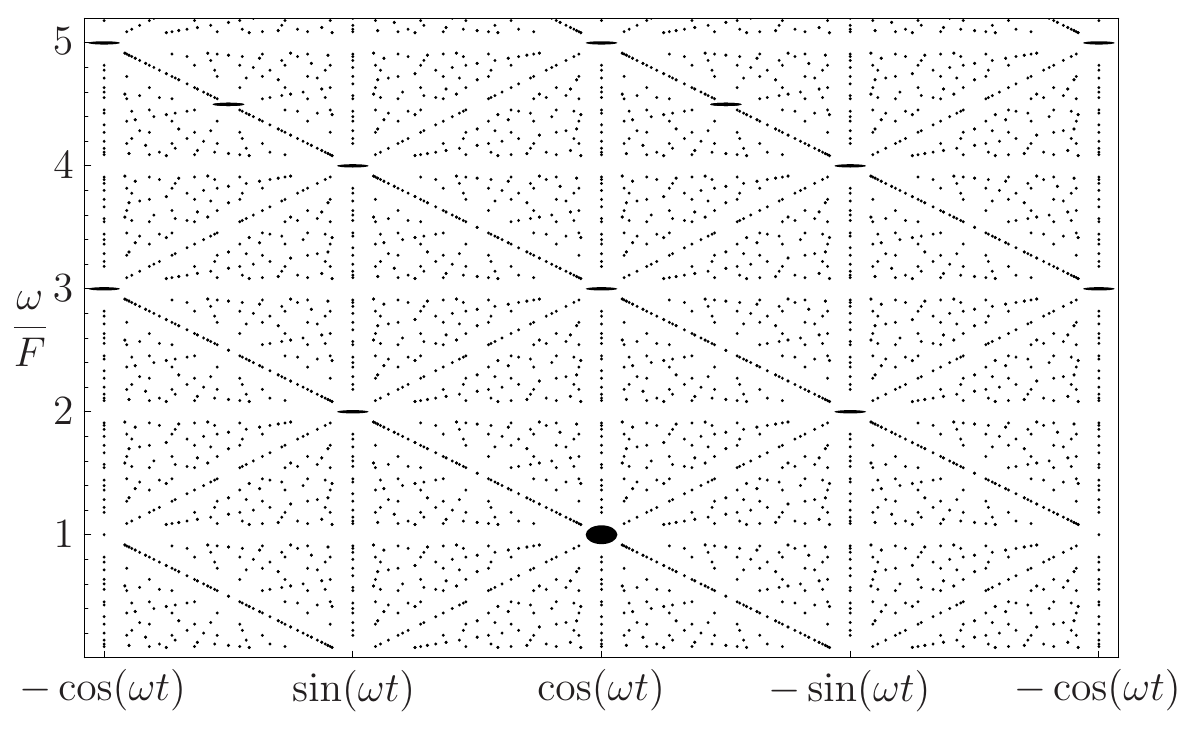}
\caption{BOs at commensurate harmonic modulation $g(t) = g_0 \cos(\omega t + \delta)$.
The stable cases $(\delta_0,\omega_0)$, according to Eq.~\eqref{stableg.eq}, are marked in the $\delta$-$\omega$ plane.
Frequencies $\omega = {l}F/{\nu}$ have been accounted for up to $\nu = 12$.
The size of the ellipses represents the life time in presence of a perturbation in phase, $\delta=\delta_0 + 10^{-4}{\rm rad}$, and frequency, $\omega=\omega_0(1+10^{-4})$.
The largest radii correspond to a life time of $5T_\text{B}$ or more, the smallest to $1T_\text{B}$ or less. 
Parameters are $F=0.2$ and $n_0 g_0 = 1$.\label{figMap}}
\end{figure}

Specifically, a modulation $g(t) = g_0 \cos(\omega t + \delta)$ at any
frequency $\omega$ commensurate with the Bloch frequency $F$ allows for
periodic BOs, if the relative phase $\delta$ is
adjusted properly [Eq.~\eqref{stableg.eq}]. In order to assess how precise this phase needs to
be adjusted, we can study the growth of fluctuations with the help of
Floquet theory by computing the life time $(\max_k \lambda_k)^{-1}$ for a certain, small phase shift of order $10^{-4}$.    
For completeness, we also consider similarly small frequency detunings, although
Floquet theory does not work because commensurability is broken, such
that we need to integrate the system of equations \eqref{dotsk} and \eqref{dotdk} numerically.
The life time is used as radius in the graphical 
representation in the $\delta$-$\omega$ plane shown in Fig.~\ref{figMap}.
The stable points are arranged on a regular pattern, where the most robust points are arranged on lines.
With increasing denominator $\nu$, the robustness drops rapidly.

%-------------------------------------------------------------------------
%-------------------------------------------------------------------------

We observe a remarkable asymmetry of the robustness between $+{\cos
  Ft}$ and $-{\cos Ft}$. The stability of the $+{\cos
  Ft}$ modulation is much more robust than the stability of  $-{\cos Ft}$. 
In the latter case, mass and interaction always have opposite sign and
the quasistatic frequencies $\omega_k$ are imaginary for $k <
k_*$. Thus, the perturbations grow and decay exponentially, and their
periodic return to the initial values relies solely on the cyclic-time
argument.  
Due to this exponential growth, the case $-{\cos F t}$ is much more
susceptible 
to perturbations than the case of $+{\cos F t}$, which has only real frequencies.
In the following, we will investigate this argument more 
quantitatively.

\subsubsection{Perturbative Lyapunov exponents}\label{secPertLya}
We consider the Floquet problem \eqref{dotsk} and \eqref{dotdk} and search for analytical solutions at a given modulation 
$g(t)=g^{(0)}(t)+g^{(1)}(t)$.
The component $g^{(0)}(t) = g_0 \cos F t$ does not destabilize the periodic time evolution, while $g^{(1)}(t)$ is a perturbation term, which may contain several frequencies and phases.
The unperturbed problem is conveniently solved using the cyclic time \eqref{eta}, which eliminates the time dependence $\cos F t$.
Then, the Bogoliubov transformation \eqref{eqBgTrafo},
with Eq.\  \eqref{eqBgDispersion} and $2m_0 = 1$,
yields the solution $c_k(\eta) \propto \exp({-i \EBg \eta})$.
It turns out to be handy to write the perturbed equation of motion for  
the excitation amplitude $\gamma_k = (c_k+c_{-k})/2$ with even parity  
(choosing the odd parity $-i(c_k-c_{-k})/2$ yields the same result below):   
\begin{align}
 i \dot \gamma_k = \cos(F t) \EBg \gamma_k + \frac{\Efr}{\EBg} n_0 g^{(1)}(t) \left( \gamma_k + \gamma_k^* \right) \, .
\end{align} 
So far, we have not made any approximation. Under the assumption that
the perturbation only causes a weak growth of $\gamma_k$ per period
$T$, we now make the ansatz 
$\gamma_k(t) = [\gamma^{0}_k + \gamma^1_k(t)] \, \exp\bigl(-i\EBg \eta(t) \bigr)$
for the first-order correction $\gamma^1_k(t)$.
In order to obtain the growth per total period $T$ of the excitation $\gamma_k$, we need to determine
\begin{align}\label{eqPertLya}
 \frac{\gamma^1_k(T)}{\gamma^0_k}
&= \frac{n_0 \Efr}{i \EBg}
\int_0^T {\rm d}t \, g^{(1)}(t) \left( 1 + \frac{\gamma^{0*}_k}{\gamma^0_k} e^{2 i \EBg \sin(F t)/F} \right)\, . 
\end{align}
Within the present approximation of a homogeneous background, the constant in the brackets of Eq.~\eqref{eqPertLya} contributes only via the zero-frequency component of $g^{(1)}$, 
causing a mere phase shift $\gamma^1_k \propto i \gamma^0_k$, which may be dropped within the leading order.

We now expand the perturbation in its frequency components
\begin{align}\label{eqNoise}
 g^{(1)}(t) &= \sum_{\nu, l} \left[ g_{\nu l} \cos\bigl(l \, \tau_{\nu 0}(t) \bigr)
 + \tilde g_{\nu l} \sin\bigl(l \, \tau_{\nu 0}(t) \bigr)\right]\ , 
\end{align}
with $\tau_{\nu 0}(t)$ from Eq.\ \eqref{deftau}, and $\nu,l > 0$ and coprime.
Within first-order perturbation theory, we may treat all contributions separately.
In accordance with Eq.~\eqref{stableg.eq}, the components $\tilde g_{\nu l}$ have vanishing contribution to the growth of excitations.
The cosine contributions, integrated according to Eq.~\eqref{eqPertLya} over a fundamental period $T=\nu \TB$, vanish for $\nu \neq 1$ and can be expressed in terms of Bessel functions $J_l$ of the first kind for $\nu=1$:
\begin{align}\label{eqRelGrowth}
 \frac{\gamma^1_k(T)}{\gamma^0_k}
&= \frac{\gamma^{0*}_k}{i \gamma^0_k} \frac{n_0 \Efr}{\EBg} \frac{2\pi}{F}
 \sum_l i^l \, g_{1l} \, J_l\left({2\EBg}/{F} \right)\ .
\end{align}
The relative growth $ \bigl| 1 +  \gamma^{(1)}_k(T)/\gamma^0_k \bigr|$ depends on the complex phase $\gamma^0_k/(\gamma^0_k)^* = e^{i \alpha}$.
For the Lyapunov exponent, we are only interested in the fastest possible growth, 
i.e., we maximize with respect to $\alpha$.
From $e^{\lambda_k T} \approx 1 + \lambda_k T = 1 + \max_{\alpha} \Re [\gamma^{(1)}_k(T)/\gamma^0_k]$ we then find
\begin{align}\label{eqLyapunovAnalytical2}
 \lambda_k = \frac{n_0 \Efr}{\EBg} 
 \biggl| \sum_{l} i^l \, g_{1 l} 
 \, J_l\left({2\EBg}/{F} \right)\biggr|\ .
\end{align}
Within the first order considered here, the growth of the excitations is caused only by the components $\nu=1$ of Eq.~\eqref{eqNoise}, i.e., integer multiples of the Bloch frequency.
Other components $\nu > 1$ not fulfilling the cyclic-time condition \eqref{eqStable2} still cause growth of perturbations, but only as a second-order effect in the small parameter $g$.
The different contributions in Eq.~\eqref{eqLyapunovAnalytical2} may add up quite differently, depending on the amplitudes $g_{1 l}$.
Consequently, the most unstable modes may be located at different values of $k$.

\subsubsection{Off-phase perturbation}
The most prominent contribution to the Lyapunov exponent \eqref{eqLyapunovAnalytical2} is the Bloch periodic perturbation $\sin(F t)$, as discussed in Ref.~\cite{Diaz2010}. The Lyapunov exponent in the case of a Bloch periodic interaction 
$g(t) = g_0 \cos(F t) + g_1 \sin(F t)$ reads
\begin{align}\label{eqLyapunovAnalytical}
 \lambda_k = \left|  g_1  n_0\, \frac{\Efr}{\EBg} \, J_1\left(\frac{2\EBg}{F} \right) \right| \, ,
\end{align}
with $\omega_k = \sqrt{k^2 (k^2 + 2 g_0 n_0)}$ [Eq.\ \eqref{eqBgDispersion} with $2 m_0=1$].
Here, we can connect to the phase perturbation of the modulation $g(t) = \pm g_0 \cos F t$, shown in the stability map of Fig.~\ref{figMap}.
We need to set $g_0 n_0 = \pm 1$ and $g_1 n_0 = 10^{-4}$.
In the case $g_0 n_0 = +1$, the Bessel function in Eq.\ \eqref{eqLyapunovAnalytical} oscillates rapidly as function of $k$, and the maximum Lyapunov exponent is found close to the sixth extremum of the Bessel function at $k \approx 1.03$ with the value $\lambda^{+}_{*} \approx 0.00035 \TB^{-1}$. 
In the case $g_0 n_0 = -1$, $\omega_k$ becomes imaginary for $k<k_* = \sqrt{2}$, while the analytic continuation of $\omega_k^{-1}$
times the Bessel function remains real and regular. And indeed, 
the maximum Lyapunov exponent is found in the region of imaginary frequencies, at $k \approx 1.05$.
There, the Lyapunov exponent  $\lambda^{-}_{*} \approx 8.84 \TB^{-1}$ is much larger than $\lambda^{+}_*$, which explains the asymmetry already observed in Fig.~\ref{figMap}.
Put differently, we find that the $+{\cos}$ modulation enhances the robustness against the sine perturbation, 
whereas the $-{\cos}$ modulation reduces it.

\subsubsection{Robustness against general noise}\label{ssSecEnhancedRobustness}
Let us come back to the more general noise \eqref{eqNoise} and the prediction for the perturbation growth \eqref{eqLyapunovAnalytical2}.
We address two questions:
To what extent can the robustness be enhanced in this case?
Do the predictions hold in realistic systems where the wave packet is wide but finite?

\begin{figure}
\includegraphics[angle=-90,width=0.95\linewidth]{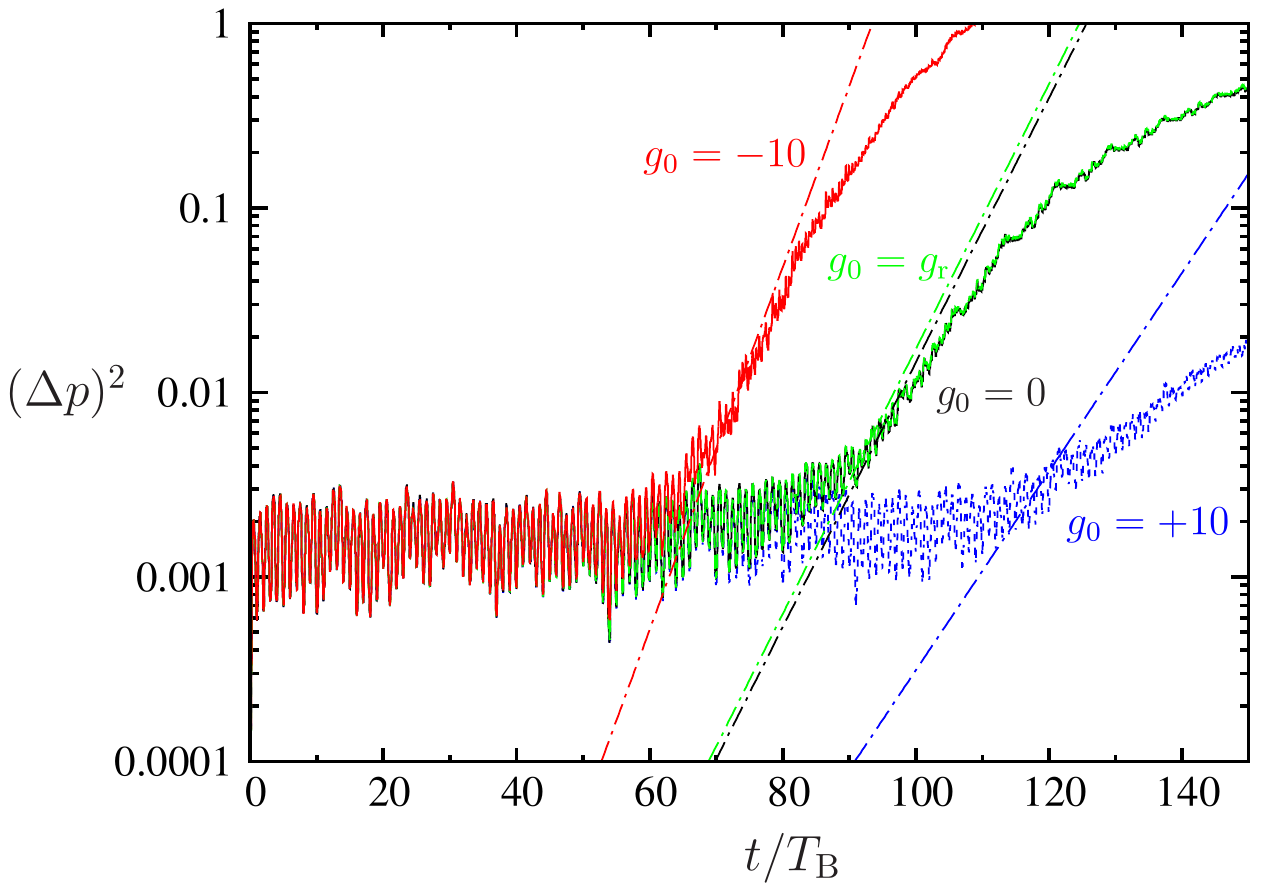}
\caption{(Color online) Momentum variance $(\Delta p)^2$ of a wave packet of soliton shape \eqref{eqSoliton.mod} with initial spatial width 
$\xi = 66.16$ (a small seed noise of $10^{-3}$ mimics experimental inhomogeneities), Bloch-oscillating in a tilted lattice with $F=0.2$.
The interaction parameter is modulated as $g(t) = g^{(1)}(t) + g_0 \cos(F t)$ for a random perturbation $g^{(1)}(t)$ of type \eqref{eqNoise} with frequencies $\Omega_l = l F /\nu$ below and above the Bloch frequency, $\nu=8$, $l=1,2,\ldots \nu^2$.
The coefficients $g_{\nu l} , \tilde g_{\nu l}$ are random numbers with zero mean and 
standard deviation 0.4, except for $g_{\nu\nu}= 0.5$ and $\tilde g_{\nu\nu} = 0$ (overwritten by $-g_0$).
The dash-dotted lines show the slope predicted by the Lyapunov exponent $\lambda = 2 \max_k \lambda_k$ [cf.\ Eq.\ \eqref{eqDelta}].
As $g_0$ is varied from $-10$ (anti\-breathing) over $g_\text{r} = -0.06$ (rigid) and $0$ (linear) [nearly overlapping with $g_\text{r}$] to $+10$ (breathing),
the predicted slope decreases and the observed life time increases.}
\label{figMomentumWidth}
\end{figure}

We thus confront the analytical result with a Gross-Pitaevskii integration [Eq.\ \eqref{eqTightBinding}], where the interaction $g(t)$ is composed of an noise term of type \eqref{eqNoise} plus a deliberate modulation $g(t) = g_0 \cos F t$.
We consider different values of $g_0$, which in the clean case define
\begin{itemize}
 \item the linear BO, $g_0=0$;
 \item the {breathing} wave packet, $g_0 >0$;
 \item the {rigid} soliton, $g_0=g_{\rm r} < 0$ [see Eq.\ \eqref{eqSoliton.mod}];
 \item the {antibreathing} wave packet, $g_0 < g_{\rm r}$.
\end{itemize}
In Fig.~\ref{figMomentumWidth}, the momentum variance is shown, which signals the decay of the wave packet and destruction of BOs.
In all cases, the momentum distribution starts to broaden at some time.
Thereby, the breathing wave packet shows a much longer life time than the linear and the rigid case, whereas the antibreathing wave packet lives much shorter.
As conjectured above, the $+$cos modulation indeed stabilizes the BOs against noise of the interaction parameter $g(t)$.

%-------------------------------------------------------------------------
%-------------------------------------------------------------------------

%---------------------------------------------------------------
\begin{figure}
\includegraphics[angle=-90,width=\linewidth]{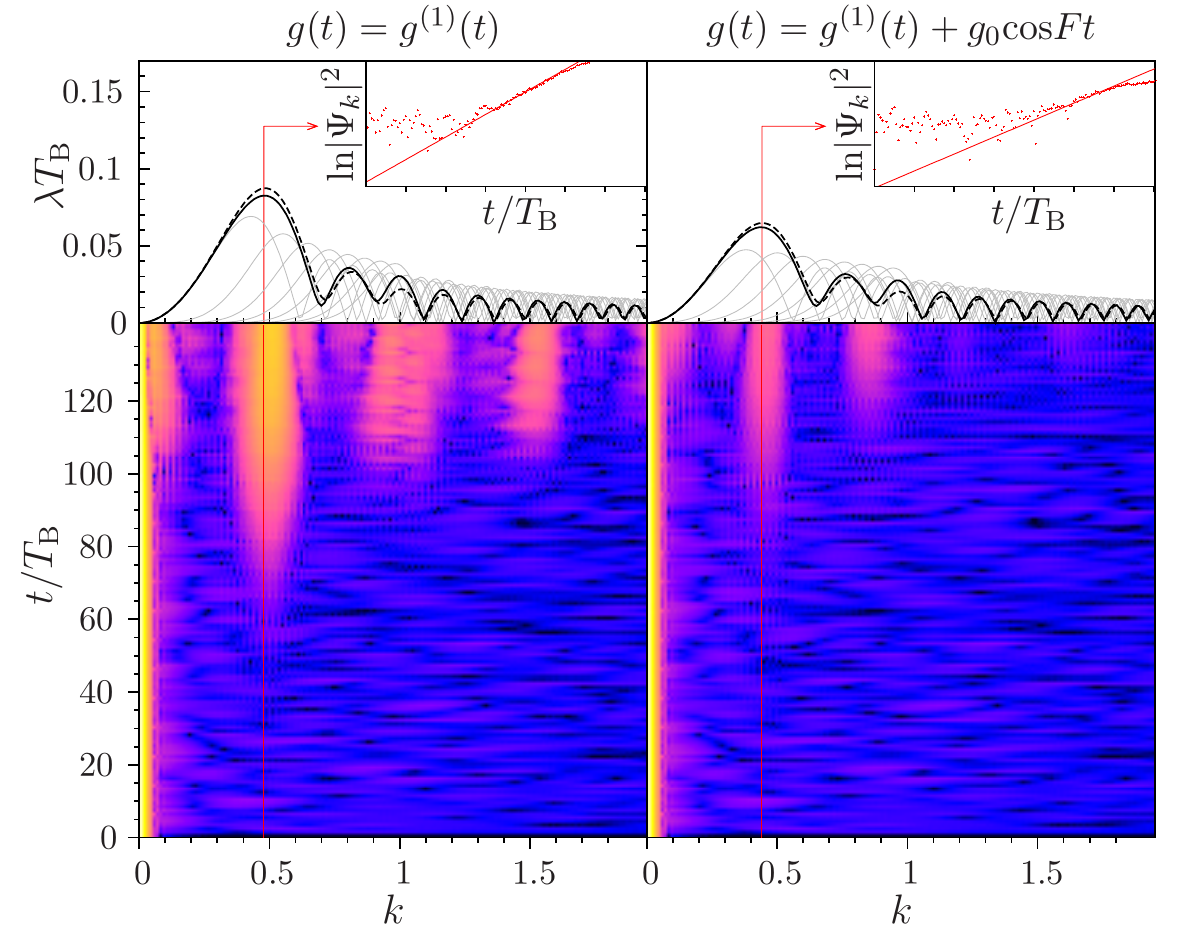}
\caption{(Color online) Momentum-space portrait of the decaying BOs in presence of a particular realization of the 
random perturbation $g^{(1)}(t)$ of type~\eqref{eqNoise},
without (left) and with (right) an additional cosine-like modulation ($g_0=10$).
Upper panel: Lyapunov exponents from Eq.\ \eqref{eqDelta} (solid black),
and the perturbative result \eqref{eqLyapunovAnalytical2} (dashed black), which is composed of the contributions from the individual frequency components 
$0.5 n_0 \Efr \left| J_l\left({2\EBg}/{F} \right)\right| / \EBg $ (thin gray).
Lower panel: Stroboscopic plot of the $k$-space density on a logarithmic color scale.
Inset: Location and growth of the most unstable mode [prediction from full Floquet theory \eqref{eqDelta}].
Parameters are the same as in
  Fig.~\ref{figMomentumWidth}. Clearly, the harmonic modulation
  stabilizes the BOs against residual fluctuations of the interaction parameter.
}\label{figKGrowth}
\end{figure}
%------------------------------------------------------------------------

In order to quantitatively verify the prediction \eqref{eqLyapunovAnalytical2}, we examine the momentum density of the wave packets, as shown in Fig.~\ref{figKGrowth} for the non-modulated and the breathing wave packet.
We can compare the growth obtained from Eq.\ \eqref{eqDelta} and the analytical prediction \eqref{eqLyapunovAnalytical2} to the growth obtained by integrating the full Gross-Pitaevskii equation \eqref{eqTightBinding}.  
Indeed, the growth of the dominantly growing mode (marked with a vertical line in Fig.\ \ref{figKGrowth}) agrees with the largest predicted Lyapunov exponents.
Beyond that, the full momentum-space distribution shows the finite width of the central wave packet and higher harmonics of the dominant mode, due to the nonlinearity of the Gross-Pitaevskii equation \eqref{eqTightBinding}.

For the momentum broadening (Fig.\ \ref{figMomentumWidth}), the accurate description of the fastest growing mode is sufficient.
Thus, Eq.\ \eqref{eqLyapunovAnalytical2} gives a satisfactory description of the BO decay.
In conclusion, we have confirmed our prediction from above, that the $+{\cos}$ modulation of $g(t)$ significantly stabilizes the BO.

\subsection{Range of validity}
The linear stability analysis presented in this section is based on the infinitely extended wave packet, and thus can only be expected to work for rather wide wave packets, as in the example of
Figs.\ \ref{figMomentumWidth} and \ref{figKGrowth}. 
In other words, the assumption \emph{wide wave packet} means that the excitations are well separated (in $k$ space) from the main wave packet. 
 
\begin{figure}[btp]
\centerline{\includegraphics[angle=270,width=\linewidth]{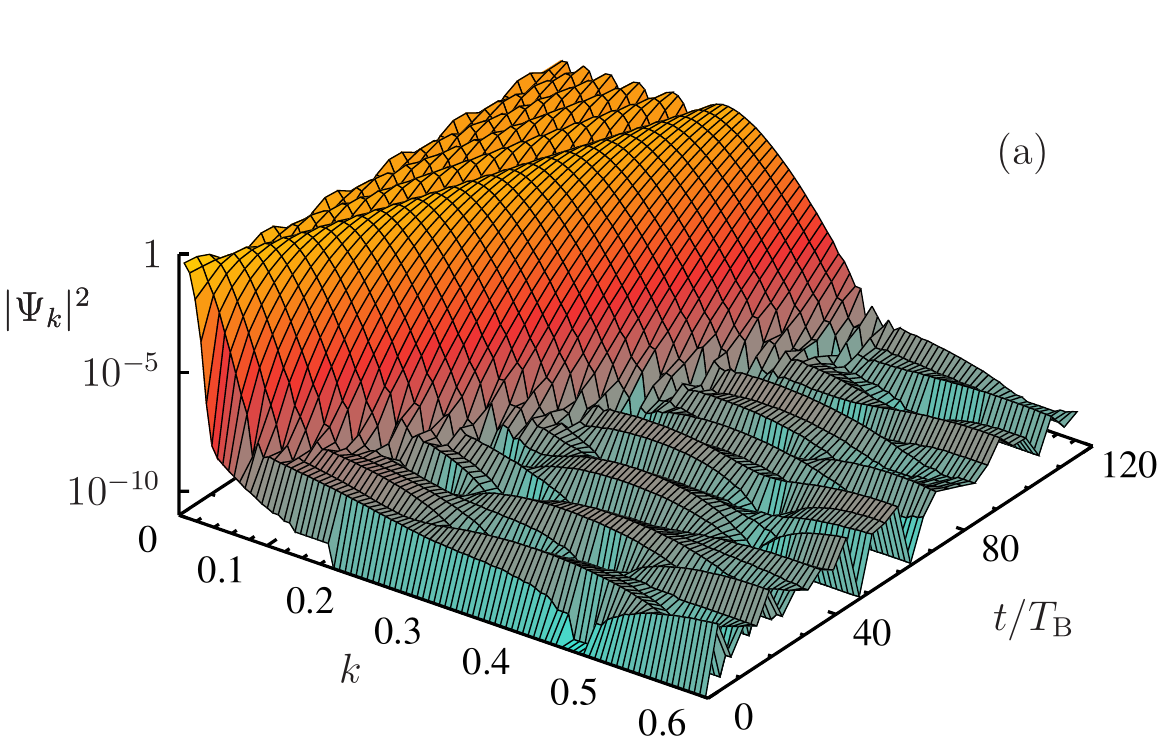}}
\centerline{\includegraphics[angle=270,width=\linewidth]{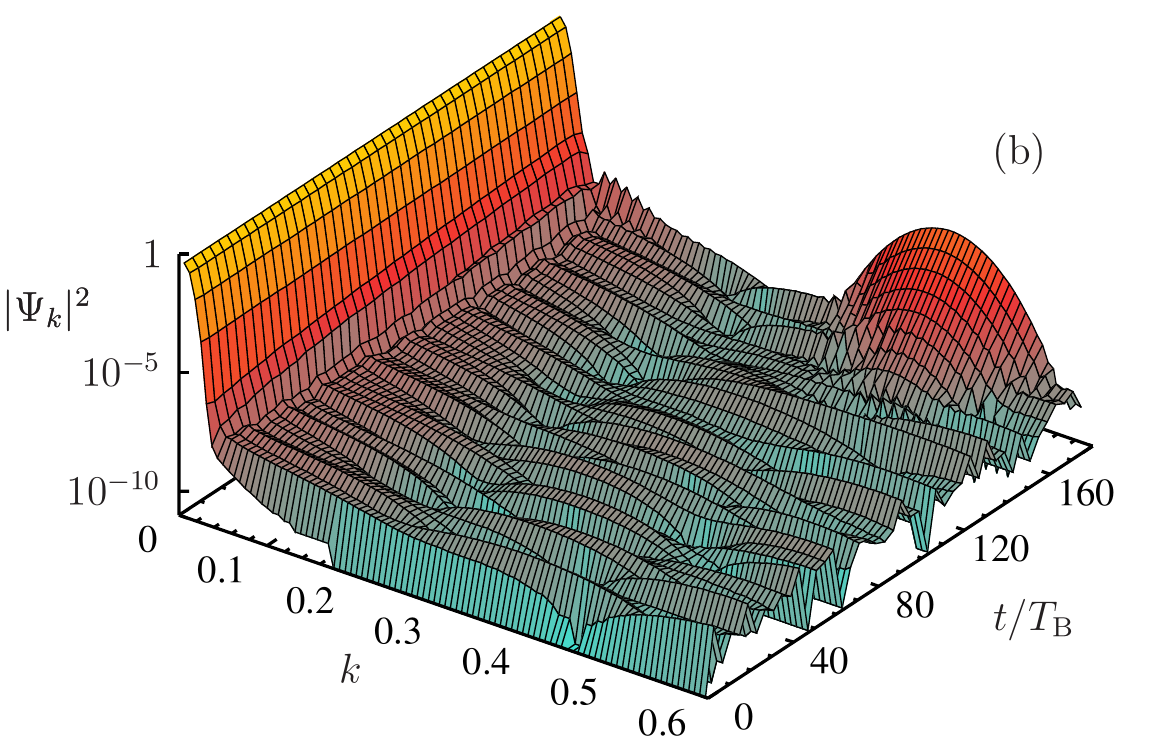}}
\caption{(Color online) Stroboscopic plot of the momentum density $|\Psi_k|^2$ in unstable cases.
(a) $g(t) = g_0$.
The immediate broadening of the wave packet in $k$-space is not compatible with the assumptions of our 
linear stability analysis, whose starting point is an infinitely extended wave function in real space.
(b)  $g(t) = g_0 \sin(F t)$.
The most unstable mode, which dominates the decay, is well separated from the central wave packet.
In both cases, the original wave function is centered around $k=0$ with width $\sigma_0=100$, $F=0.2$ and $g_0=2.5$.
}\label{figKSpacePortraits}
\end{figure}

Interactions $g(t)$ with non-zero time average, like $g(t) = g_0$, act directly on the width degree of freedom, as discussed in Sec.~\ref{sSecCCInteraction}. 
In this case, the homogeneous stability analysis is not reliable, because the most important degree of freedom
is missed. 
Indeed, the momentum width $\Delta p = \sqrt{r}$ increases linearly with time, as pointed out in subsection~\ref{sSecCCInteraction}, Eq.\ \eqref{r_g0}.
Figure \ref{figKSpacePortraits} (a) shows exactly this:
In the case of constant nonlinearity, the central wave packet is spreading from the start and soon covers a large range in momentum space, such that one cannot consider excitations separated from the central wave packet.
In this case, the dynamics is better described by  the collective coordinates approach of Sec.~\ref{secCollectiveCoordinates}.
Both approaches must be considered as complementary to each other to describe a wide range of situations.

%----------------------------------------------------------
%----------------------------------------------------------

On the contrary, Fig.~\ref{figKSpacePortraits} (b) shows that in the case of the sine perturbation, the perturbations remain well separated from the central wave packet.
In this case, and in the other cases with vanishing mean of $g(t)$, the linear stability analysis proves to be very powerful.

%----------------------------------------------------
\section{Conclusions}
We have treated the problem of BOs with a time-dependent interaction in the
mean-field framework of the one-dimensional tight-binding model.
This description applies to a dilute Bose gas in a deep lattice
potential with a strong transverse confinement, as well as to arrays of nonlinear
optical wave-guides.
Interestingly, the stability of BOs in presence of modulated
interactions is sensitively conditioned on the relative phase between
modulation and BO. 

Our analysis shows that already the linear BO has a breathing width, but its momentum-space
distribution is time-independent (up to the uniform translation
$p=-Ft$). 
For sufficiently wide wave packets, a cyclic-time argument allows
identifying a class of interactions $g(t)$ that lead to periodic
dynamics, in spite of the interaction. 
In these cases, both real-space and momentum-space distribution become time dependent, but return periodically to the initial state.
In all other cases, the BO decays, with simultaneous momentum-space broadening.
The broadening is either due to the broadening of the central ($k$-space) wave packet, or due to the growth of excitations separated from the central wave packet. 

In order to quantitatively describe both the periodic cases and the decay, we have employed two complementary methods, the \emph{collective-coordinates} approach and the \emph{linear stability analysis} of the extended wave packet.
The collective-coordinates approach is valid as long as the shape of the wave packet is essentially conserved.
It is capable of describing, on the one hand, the centroid and the breathing dynamics in the periodic cases and, on the other hand, the beginning of the decay in the unstable cases; for example, at constant interaction.
The linear stability analysis of the extended wave packet is suitable for the quantitative description of the decay of wide wave packets, when the relevant excitations are well decoupled in $k$-space from the original wave packet.
Together, the two approaches provide a rather complete picture of the wave-packet
dynamics.

The most striking prediction due to the linear stability analysis is that a cosine modulation of the interaction (the one that enhances the breathing) makes the BO more robust with respect to certain perturbations.
This works especially well for a fluctuating residual interaction with zero time average, as in Figs.\ \ref{figMomentumWidth} and \ref{figKGrowth}. 
However, the modulation has little effect on the decay due to a finite offset of the interaction. The strategy for achieving long-living BOs is to tune the interaction to zero as well as possible and then employ the cosine modulation to minimize the effect of residual fluctuations around zero.

Conversely, we conjecture that the enhanced phase sensitivity of nonlinear BOs (similar to the effect of harmonically shaken lattices in the linear case \cite{Creffield2011}), may become useful to design high-precision matter-wave interferometers. 
 \label{secConclusions}

\begin{acknowledgments} 
This work was supported in Madrid by Ministerio de Ciencia e Innovaci\'{o}n (MICINN)
(projects MOSAICO and MAT2010-17180) and in Singapore by the National Research Foundation \& Ministry of Education.  
C.~G. acknowledges funding by CEI Campus Moncloa, PICATA program.
R. P. A. L. acknowledges financial support by Conselho Nacional de Desenvolvimento Cient\'{\i}fico e Tecnol\'ogico (CNPq).
C.~G. and C.~A.~M. acknowledge financial support by Deutsche For\-schungs\-ge\-mein\-schaft (DFG) for the time when both were affiliated with Universit\"{a}t Bayreuth. 
Travel between Bayreuth and Madrid was supported by the joint program Acciones Integradas of Deutscher Akademischer Auslandsdienst (DAAD) and MICINN.
\end{acknowledgments}

\bibliography{references}

\end{document}